\documentclass[12pt,a4]{article}

\newcommand{\be}{\begin{equation}}
\newcommand{\ee}{\end{equation}}
\newcommand{\bee}{\begin{eqnarray}}
\newcommand{\beee}{\begin{array}}
\newcommand{\eee}{\end{eqnarray}}
\newcommand{\eeee}{\end{array}}
\newcommand{\ga}{\alpha}

\newcommand{\ie}{{\it i.e.,} }

\newcommand{\q}{\,,\qquad}

\usepackage{amsthm,amsmath,latexsym,multibox,amssymb,amsfonts}
\usepackage{graphicx,lscape,fancyhdr,array,stmaryrd}

\newcommand{\bep}{\begin{picture}}
\newcommand{\eep}{\end{picture}}

\newcounter{YoungHeight}\newcounter{YoungWidth}

\newcounter{Mul1}\newcounter{Mul2}\newcounter{Mul3}\newcounter{Mul4}
\newcounter{A0}\newcounter{A1}\newcounter{A2}\newcounter{A3}\newcounter{A4}\newcounter{A5}\newcounter{A6}
\newcounter{B0}\newcounter{B1}\newcounter{B2}\newcounter{B3}
\newcounter{C1}\newcounter{C2}\newcounter{C3}\newcounter{C4}\newcounter{C6}\newcounter{C7}
\newcounter{D1}\newcounter{D2}\newcounter{D3}\newcounter{D4}\newcounter{D5}
\newcounter{T0}\newcounter{T1}

\newcounter{TGR0}
\newcounter{R0}\newcounter{R1}\newcounter{R2}\newcounter{R3}
\newcounter{AR0}\newcounter{AR1}\newcounter{AR2}\newcounter{AR3}\newcounter{AR5}
\newcounter{Dotted0}\newcounter{Dotted1}\newcounter{Dotted2}\newcounter{Dotted3}

\newcounter{reptA}
\newlength{\txtHShift}

\newlength{\txtWidth}

\newcommand{\HalfLength}[2]{\setcounter{Mul1}{#1}\setcounter{Mul2}{#1}\addtocounter{Mul1}{\value{Mul2}}\addtocounter{Mul1}{\value{Mul2}}%
\addtocounter{Mul1}{\value{Mul2}}\addtocounter{Mul1}{\value{Mul2}}\setcounter{#2}{\value{Mul1}}}

\newcommand{\Add}[3]{\setcounter{#1}{#2}\addtocounter{#1}{#3}}

\newcommand{\Length}[1]{#10}
\newcommand{\YoungScale}{}%\unitlength=0.35mm}

\newcommand{\shiftedText}[2]{{\hspace{#1}#2}}
\newcommand{\calcHShift}[1]{\settowidth{\txtWidth}{#1}\setlength{\txtHShift}{-0.5\txtWidth}}

\newcommand{\TextTop}[3]{{\calcHShift{#1}\HalfLength{#2}{T0}\Add{T1}{\Length{#3}}{-9}\put(\value{T0},\value{T1}){\shiftedText{\txtHShift}{#1}}}}

\newcommand{\BlockA}[2]{{\YoungScale\bep(\Length{#1},\Length{#2}){\Add{A1}{#1}{1}\Add{A2}{#2}{1}}%
\multiput(0,0)(10,0){\value{A1}}{\line(0,1){\Length{#2}}}\multiput(0,0)(0,10){\value{A2}}{\line(1,0){\Length{#1}}}%
\setcounter{YoungHeight}{\Length{#2}}\setcounter{YoungWidth}{\Length{#1}}\eep}}

\newcommand{\BlockB}[4]{{\YoungScale\Add{B3}{\Length{#2}}{\Length{#4}}%
\bep(\Length{#1},\value{B3})\put(0,\Length{#4}){\BlockA{#1}{#2}}%
\put(0,0){\BlockA{#3}{#4}}\setcounter{YoungHeight}{\value{B3}}\setcounter{YoungWidth}{\Length{#1}}\eep}}

\newcommand{\RectT}[3]{\bep(\Length{#1},\Length{#2})\put(0,0){\line(1,0){\Length{#1}}}\put(0,0){\line(0,1){\Length{#2}}}%
\put(\Length{#1},\Length{#2}){\line(-1,0){\Length{#1}}}\put(\Length{#1},\Length{#2}){\line(0,-1){\Length{#2}}}#3{#1}{#2}\eep}

\newcommand{\RectARow}[2]{{\bep(\Length{#1},10)\put(0,0){\RectT{#1}{1}{\TextTop{#2}}}\eep}}

\newcommand{\RectBRow}[4]{{\bep(\Length{#1},20)\put(0,0){\RectT{#2}{1}{\TextTop{#4}}}%
\put(0,10){\RectT{#1}{1}{\TextTop{#3}}}\eep}}

\newcommand{\RectCRow}[6]{{\bep(\Length{#1},30)\put(0,0){\RectT{#3}{1}{\TextTop{#6}}}%
\put(0,10){\RectT{#2}{1}{\TextTop{#5}}}\put(0,20){\RectT{#1}{1}{\TextTop{#4}}}\eep}}

\newcommand{\RectBYoung}[3]{{\bep(0,0)\put(0,0){#3}\eep\Add{A0}{\value{YoungHeight}}{10}%
\bep(\Length{#1},\value{A0})%
\put(0,\value{YoungHeight}){\RectT{#1}{1}{\TextTop{#2}}}\eep}}

\newcommand{\YoungA}{\BlockA{1}{1}}
\newcommand{\YoungB}{\BlockA{2}{1}}

\newcommand{\YoungBA}{\BlockB{2}{1}{1}{1}}

\newcommand{\YoungAAA}{\BlockA{1}{3}}

\usepackage{amsthm,amsmath,latexsym,multibox,amssymb,amsfonts}
\usepackage{graphicx,lscape,fancyhdr,array,stmaryrd,euscript,wrapfig}

\begin{document}
\begin{titlepage}
$$$$

\vspace{2cm}

\begin{flushright}
 FIAN/TD/29-09\\
\vspace{-1mm}
%{XXXX 1996}\\
\end{flushright}\vspace{1cm}

\begin{center}
{\bf \Large Frame-Like Action and Unfolded Formulation for Massive
Higher-Spin Fields} \vspace{1cm}

\textsc{D.S. Ponomarev and M.A.
Vasiliev}

\vspace{.7cm}

{ I.E.Tamm Department of Theoretical Physics, P.N.Lebedev Physical
Institute,\\Leninsky prospect 53, 119991, Moscow, Russia\\
ponomarev@lpi.ru\,,\quad vasiliev@lpi.ru }

\end{center}

\vspace{0.5cm}
\begin{abstract}
Unfolded equations of motion for symmetric massive bosonic fields
of any spin in Minkowski and $(A)dS$ spaces are presented.
Manifestly gauge invariant action for a spin $s\geq 2$ massive field
in any dimension  is constructed in terms of gauge invariant curvatures.
\end{abstract}

\end{titlepage}

\tableofcontents

\numberwithin{equation}{section}

\section{Introduction}

Although elementary particles of spins higher than two are unlikely
to be directly observed in the modern high-energy experiments, there
are several reasons to believe that they play a fundamental role at
ultra high energies relevant to quantum gravity. Indeed, infinite
towers of massive higher-spin (HS) fields are important for
consistency of String Theory \cite{str1, str2, str3}. On the other
hand, existence of nonlinear theories of massless HS fields with
unbroken HS symmetries indicates new remarkable structures
underlying a theory of fundamental interactions (for recent reviews
on HS theory see, e.g., \cite{Vasilievobzor, Sorokin, Sundellobzor,
SolvayWorkshop, FotopoulosTsulaia, Iazeolla}). To uncover relation
between the two types of theories and, more generally, to understand
a spontaneous breakdown mechanism for HS symmetries that gives rise
to a theory of HS massive fields, it is important to extend the
approach, which underlies nonlinear massless HS theories, to massive
HS fields starting from the linearized level. This is the goal of
this paper.

There are two  approaches to HS fields.
The metric-like approach
 \cite{SinghHagen, Fronsdal, dewitfreedman, ZinovievMetric, unconstrainted,
 Metsaev,Buchbinder} generalizes metric formulation of gravity.
 The frame-like formalism \cite{Vasiliev1980, Vasiliev1987, LopatinVasiliev, Vasiliev2001, SkvortsovPM,
 ZinovievFrame}, that generalizes Cartan formulation of gravity,
  deals with differential forms.
It reproduces the metric-like formulation in a particular gauge and
is most appropriate to control gauge HS symmetries and interactions
\cite{Fradkin:1987ks,Vasiliev1990, Vasiliev2003}. The study of the frame-like
approach in the context of massive HS fields was initiated by
Zinoviev  who constructed gauge invariant actions in terms of the
frame-like fields in \cite{ZinovievFrame} where, however, the gauge
invariance of the action was not manifest, requiring the adjustment
of coefficients in the action from the condition of gauge invariance.
In this paper we propose the manifestly gauge invariant action for
symmetric
 massive HS fields formulated in terms of gauge invariant HS curvatures.
The same time, these curvatures underly the construction of
unfolded formulation of free massive HS field equations which is also given
in this paper.

Free equations for symmetric massive fields of any spin
are\footnote{  For notation see Appendix A} \cite{Dirac}
\begin{equation}
\label{metriceq}
(\partial^n\partial_n+m^2)\phi_{a(s)}=0\q
\partial^n \phi_{na(s-1)}=0\q
\phi^{n}{}_{n a(s-2)}=0\,.
\end{equation}
Our aim is to reformulate the equations (\ref{metriceq})
 in the unfolded form  of generalized zero-curvature equations
 \cite{Vasiliev1988unf}
 (for a review see \cite{SolvayWorkshop})
\begin{equation}
\label{2.1}
R^{\alpha}(x)\stackrel{def}{=}dW^{\alpha}(x)+G^{\alpha}(W(x))=0,
\end{equation}
where $d=dx^{\mu}\partial_{\mu}$ is the exterior differential and
\begin{equation}
\label{2.2}
G^{\alpha}(W^{\beta})\stackrel{def}{=}\sum_{n=1}^{\infty}f^{\alpha}_{\beta_1 \dots \beta_n}W^{\beta_1}\dots W^{\beta_n}
\end{equation}
is built from the exterior product (which is implicit in this section)
 of differential forms
$W^{\beta}(x)$ and satisfies the compatibility condition
\begin{equation}
\label{2.3}
G^{\beta}(W)\frac{\delta^L G^{\alpha}(W)}{\delta W^{\beta}}\equiv 0\,.
\end{equation}
Here index $\alpha$ enumerates a set
of differential forms, which may be infinite.
%(in this paper fields are valued in
%Young diagrams \cite{Fulton, SolvayWorkshop}).

Unfolding consists of the equivalent reformulation of a system of partial
differential equations in the form (\ref{2.1}), which is always possible by
virtue of introducing enough (may be infinite towers of) auxiliary
and Stueckelberg fields. The fields, that neither can be expressed in terms of derivatives of other
fields, nor can be gauged away are called {\it dynamical}.
{\it Auxiliary fields} are expressed via derivatives
of all orders of the dynamical field $\phi_{a(s)}$. {\it Stueckelberg fields}
are pure gauge with respect to algebraic shift symmetries.
Differential conditions
 on the dynamical fields imposed by the unfolded equations are called
{\it dynamical equations}.  (Note that the standard tool for the analysis of
physical content of unfolded equations is provided by the $\sigma_-$-cohomology technics
\cite{Shayns0, SolvayWorkshop}.)

Unfolded field equations (\ref{2.1}) are manifestly invariant under the gauge transformation
with a degree $p^\alpha -1$ differential form gauge parameter
$\varepsilon^\alpha(x)$ associated to any degree $p^\alpha >0$ form $W^\alpha$
\begin{equation}
\label{2.6}
\delta W^{\alpha}=d\varepsilon^{\alpha}-
\varepsilon^{\beta}\frac{\delta^LG^{\alpha}(W)}{\delta W^{\beta}}\,
\end{equation}
because
\begin {equation}
\delta R^{\alpha} =-R^{\gamma}\frac{\delta^L}{\delta W^{\gamma}}\left(\varepsilon^{\beta}\frac{\delta^L G^{\alpha}(W)}{\delta W^{\beta}}\right)\,.
\end{equation}
The gauge transformations of 0-forms only contain the gauge
parameters $\varepsilon^{\alpha_1}$ associated to 1-forms $W^{\alpha_1}$ in (\ref{2.6}):
\begin{equation}
\label{2.7}
\delta C^{\alpha_0}=-\varepsilon^{\alpha_1}\frac{\delta^LG^{\alpha_0}(W)}{\delta W^{\alpha_1}}\,.
\end{equation}

In the unfolded dynamics approach,
background Minkowski or $(A)dS$ geometry is described by the 1-form frame field
  (vielbein) $e^a=e^a_{\mu}dx^{\mu}$ and Lorentz spin-connection 1-form
  $\omega^{a,b}=\omega_{\mu}^{a,b}dx^{\mu}$, $\omega^{a,b}=-\omega^{b,a}$, that
  obey the equations
 \begin{gather}
 \label{Torsion}
 T^a\equiv de^a+\omega^{a,}{}_b\wedge e^b=0,\\
 \label{Curvature}
 R^{a,b}\equiv d\omega^{a,b}+\omega^{a,}{}_c\wedge \omega^{c,b}-\lambda^2
 e^a\wedge e^b=0\,,
 \end{gather}
 where $-\lambda^2$ is the cosmological constant.
 The vielbein $e_\mu^a$ is nondegenerate and relates
 fiber (tangent) and  base (world) indices. Eq.~(\ref{Torsion}) is
 the zero-torsion condition that expresses the Lorentz connection
 $\omega^{a,b}$ via vielbein $e^a$. The equation (\ref{Curvature}) then describes the $AdS_d$ space with the symmetry algebra $o(d-1,2)$ for $\lambda^2>0$, $dS_d$ space with the symmetry algebra $o(d,1)$ for $\lambda^2<0$ and  Minkowski space for $\lambda^2=0$.

Unfolded formulation  brings together such
important issues as coordinate independence due to using the exterior
algebra formalism and manifest gauge invariance (Stueckelberg in the massive
case) that, controls a number of degrees of freedom in the system.  The
unfolded dynamics
approach made it possible to find a full nonlinear system of equations for massless
fields in four \cite{Vasiliev1990} and any \cite{Vasiliev2003} dimensions.
Extension of the unfolded description to the massive case
 should shed light on the mechanism of  spontaneous breakdown
of HS gauge symmetries and, eventually, unfolded formulation of String
Theory to establish a  correspondence
between the latter and HS theory.

Our aim is to find an explicit form of the unfolded system of equations that
contains a dynamical equation (\ref{metriceq}) on the dynamical field
$\phi_{a(s)}$ plus constraints that
express auxiliary fields in terms of $\phi_{a(s)}$.
The unfolded form of massive field
equations was worked out in \cite{Shayns0} for the case of spin 0.
General features of the unfolded description of massive fields of any spin were
 analyzed in \cite{BIS}.

In this paper we present the explicit form of the unfolded equations for the
very particular case of spin $s$ massive symmetric gauge fields
in Minkowski  and $(A)dS$ space which, as we show,
can be described by a set of 1-form connections and 0-form Weyl
tensors  valued in various Lorentz tensor representations described by
two-row Young tableaux.
In contrast to massless and partially massless HS fields and in agreement with
the results of \cite{BIS},
gauge invariant curvatures for massive fields necessarily involve
0-forms. Using the gauge invariant HS curvatures that result from
the unfolded formulation of massive fields we also present
the manifestly gauge invariant actions for HS massive fields.

The layout of the rest of the paper is as follows.
 Unfolded equations, that describe Weyl module of a spin $s$ massive field,
 are derived in Section 2. Extension of these equations to 1-form gauge
 potentials is given in Section 3.
Manifestly gauge invariant action for massive HS fields is
constructed in Section 4. Section 5 contains conclusions. Our conventions
and some useful formulas are given in Appendices A and B, respectively.

\section{Forms of definite degree}
\label{samedegree}

\subsection{General form of curvatures}
The  frame-like formulation of massless \cite{Vasiliev1980, Vasiliev1987,
 LopatinVasiliev, Vasiliev2001} and partially massless \cite{SkvortsovPM}
 fields
 operates with 1-forms valued in irreducible  Lorentz tensor spaces described by
 two-row Young tableaux (for more details on Young tableaux see, e.g.,
 \cite{SolvayWorkshop, Fulton}; the facts relevant to the consideration
 of this paper are given in  Appendix A). Let $\mathbf{Y}(k,l)$ denote
 the space of tensors described by
 two-row Young tableaux with $k$ cells in the first row and $l\le k$ cells
  in the second row.
 In this section we analyze the most general linear unfolded equations
 that can be formulated in terms of $p-$forms $W^{a(k),b(l)}$
 with some definite $p$, valued in $V=\sum_{k\geq l\geq0}\oplus \mathbf{Y}(k,l)$.
 Since the analysis is insensitive to $p$, it is kept arbitrary
 in this section.

Unfolded equations  formulated for differential
forms of a definite degree, valued in $V$, read as
  \begin{equation}
  \label{eqhom}
 0=R^{a(k),b(l)}=DW^{a(k),b(l)}+G^{a(k),b(l)}(W,e)\,,
\end{equation}
 where $D$ is the Lorentz covariant derivative
\begin{equation}
\label{CovDer}
DC^{ab\dots}=dC^{ab\dots}+\omega^{a,}{}_cC^{cb\dots}+\omega^{b,}{}_cC^{ac\dots}+\dots\,.
\end{equation}
{}From (\ref{Curvature}) it follows that
\begin{equation}
\label{dsq}
D^2 C^{a(k),b(l)}=\lambda^2(ke^ae_cC^{ca(k-1),b(l)}+
 le^be_cC^{a(k),cb(l-1)}).
 \end{equation}
The background vielbein 1-form $e^a=dx^\mu e_{\mu}^a$
  enters $G^{a(k),b(l)}(W,e)$ linearly in (\ref{eqhom}).  Generally $e^a\wedge
W^{\ldots}$ is a $(p+1)$-form valued in a reducible Lorentz
module. Since vielbein carries one fiber index, the projection  to
irreducible components described by  two-row Young diagrams gives
$${  \atop\RectBRow{5}{3}{$k$}{$l$}}{  \atop\otimes}{   \atop \YoungA}{  \atop=}{\sigma_+^1 \atop\RectBRow{5}{3}{$k+1$}{$l$}}{  \atop\oplus}
{\sigma_+^2  \atop\RectBRow{5}{3}{$k$}{$l+1$}}{  \atop\oplus}{\sigma_-^1  \atop\RectBRow{5}{3}{$k-1$}{$l$}}{ \atop\oplus}
{\sigma_-^2  \atop\RectBRow{5}{3}{$k$}{$l-1$}}{  \atop+\ldots \,,}$$
where  tensors described by three-row Young tableaux are projected out.
Hence, there exist four linearly independent operators built from the
vielbein 1-form.
$\sigma_+^1$ and $\sigma_+^2$ ($\sigma_-^1$ and $\sigma_-^2$)
increase (decrease), respectively, the lengths of the first and the second
row by one.
 Operators $\sigma$ will be sometimes referred to as ``cell operators''.

Thus, the curvature (\ref{eqhom}) can be represented in the form
\begin{equation}
\label{3.2}
R(W)=DW+\sigma_+^1(W)+
\sigma_+^2(W)+  \sigma_-^1(W)+ \sigma_-^2(W)\,,
\end{equation}
where the form of the cell operators is determined up to overall
coefficients by the properties of traceless two-row Young diagrams
\begin{equation}
\label{3.5}
\;\;\qquad\qquad\sigma_-^1(k,l)(W^{a(k),b(l)})=f(k,l)
(e_mW^{a(k-1)m,b(l)}+\frac{l}{k-l+1}e_mW^{a(k-1)b,b(l-1)m})\,,
\end{equation}
\begin{equation}
\label{3.5.1}
\sigma_-^2(k,l)(W^{a(k),b(l)})=F(k,l)(e_mW^{a(k),b(l-1)m})\,,\qquad\qquad\qquad\qquad\qquad
\quad
\end{equation}
\begin{equation*}
\qquad\qquad\sigma_+^1(k,l)(W^{a(k),b(l)})=g(k,l)(k+1)(e^aW^{a(k),b(l)}-
\frac{k}{d+2k-2}e_mW^{a(k-1)m,b(l)}\eta^{aa}-\;
\end{equation*}
\begin{equation}
\label{3.5.2}
-\frac{l}{d+k+l-3}e_mW^{a(k),mb(l-1)}\eta^{ab}
+\frac{lk}{(d+2k-2)(d+k+l-3)}e_mW^{a(k-1)b,b(l-1)m}\eta^{aa})\,,
\end{equation}
\begin{equation*}
\sigma_+^2(k,l)(W^{a(k),b(l)})=G(k,l)(l+1)(e^bW^{a(k),b(l)}-\frac{k}{k-l}e^aW^{a(k-1)b,b(l)}-\quad
\end{equation*}
\begin{equation*}
-\frac{l}{d+2l-4}e_mW^{a(k),b(l-1)m}\eta^{bb}-\frac{k(k-l-1)}{(k-l)(d+k+l-3)}e_mW^{a(k-1)m,b(l)}\eta^{ab}+\qquad\quad
\end{equation*}
\begin{equation*}
+\frac{lk(d+2k-4)}{(k-l)(d+2l-4)(d+k+l-3)}e_mW^{a(k-1)b,b(l-1)m}\eta^{ab}+\qquad\qquad\qquad\qquad
\qquad
\end{equation*}
\begin{equation*}
+\frac{k(k-1)}{(k-l)(d+k+l-3)}e_mW^{a(k-2)mb,b(l)}\eta^{aa}-\qquad\qquad\qquad\qquad\qquad\qquad\qquad\qquad
\end{equation*}
\begin{equation}
\label{3.5.3}
-\frac{lk(k-1)}{(k-l)(d+k+l-3)(d+2l-4)}e_mW^{a(k-2)bb,b(l-1)m}\eta^{aa})\,.\qquad\qquad\qquad\quad
\;\:
\end{equation}

To rule out the terms with
Young diagrams $\mathbf{Y}(k,l)$ with $l>k$, that are zero, it is convenient to
demand
\begin{equation}
\label{oblopred}
f(k,l)=F(k,l)=g(k,l)=G(k,l)=0 \quad \text{for} \quad l>k
\end{equation}
and
\begin{equation}
\label{fgzero}
f(n,n)=G(n,n)=0.
\end{equation}

%\subsection{Bianchi identities}

The compatibility condition (\ref{2.3}) for the equation (\ref{3.2})
is
\begin{equation}
\label{sigmasquared}
D^2+\sigma^2=0\,, \qquad \sigma=\sigma_-^1+\sigma_-^2+\sigma_+^1+\sigma_+^2\,.
\end{equation}
Restriction of (\ref{sigmasquared}) to different types of Young tableaux
yields the following conditions
\begin{equation}
\label{3.8}
(\sigma_-^1)^2=0,\quad (\sigma_-^2)^2=0,\quad (\sigma_+^1)^2=0,\quad (\sigma_+^2)^2=0,
\end{equation}
\begin{equation}
\label{3.9}
\{\sigma_-^1,\sigma_-^2\}=0, \quad \{\sigma_-^1,\sigma_+^2\}=0, \quad \{\sigma_-^2,\sigma_+^1\}=0, \quad \{\sigma_+^1,\sigma_+^2\}=0,
\end{equation}
\begin{equation}
\label{3.10}
D^2+\{\sigma_-^1,\sigma_+^1\}+\{\sigma_-^2,\sigma_+^2\}=0\,.
\end{equation}
The conditions (\ref{3.8})  are trivially satisfied due to
the antisymmetry of the exterior product.
The conditions (\ref{3.9}) give the following constraints on the coefficients
$F(k,l)$, $G(k,l)$, $f(k,l)$ and $g(k,l)$
\begin{equation}
\label{3.11}
\begin{split}
\frac{f(k,l-1)F(k,l)}{f(k,l)F(k-1,l)}=\frac{k-l}{k-l+1}  :  \quad \{\sigma_-^1,\sigma_-^2\}=0,\\
\frac{G(k-1,l)f(k,l)}{G(k,l)f(k,l+1)}=\frac{(k-l-1)(k-l+1)(d+k+l-2)}{(k-l)(k-l)(d+k+l-3)} :\quad \{\sigma_-^1,\sigma_+^2\}=0,\\
\frac{F(k+1,l)g(k,l)}{F(k,l)g(k,l-1)}=\frac{d+k+l-3}{d+k+l-2} : \quad \{\sigma_-^2,\sigma_+^1\}=0,\\
\frac{g(k,l+1)G(k,l)}{g(k,l)G(k+1,l)}=\frac{k-l+2}{k-l+1} : \quad
\{\sigma_+^2,\sigma_+^1\} =0\,.
\end{split}
\end{equation}
Note that only three of these conditions turn out to be independent.

\subsection{Field redefinition ambiguity}
Eq. (\ref{3.2}) has  the
freedom in the field redefinition
 \begin{equation}
\label{3.15}
 W\rightarrow \tilde{W}, \qquad W^{a(k),b(l)}=\beta(k,l)\tilde{W}^{a(k),b(l)},
 \qquad \beta(k,l)\ne 0\,,
 \end{equation}
 which induces  the following redefinition of the coefficients
 \begin{equation}
 \label{3.16}
 \begin{split}
\tilde{g}(k,l)= g(k,l)\frac{\beta(k,l)}{\beta(k+1,l)},\quad
 \tilde{G}(k,l)=G(k,l)\frac{\beta(k,l)}{\beta(k,l+1)},\\
\tilde f(k,l)=f(k,l)\frac{\beta(k,l)}{\beta(k-1,l)},\quad
 \tilde F(k,l)=F(k,l)\frac{\beta(k,l)}{\beta(k,l-1)}.
\end{split}
\end{equation}
For the future convenience of the analysis of variation of the action in Section
 \ref{freeaction} we
fix this ambiguity by demanding the operators
$\sigma_+^{1,2}$ be conjugated to $\sigma_-^{1,2}$ up to some factors $n$ and $N$
with respect to the scalar product
\begin{equation}
\label{1formsgeneral1}
\langle\psi_{\{p\}}^{a(k),b(l)}|\phi_{\{q\}}^{a(k),b(l)}\rangle=
\int \epsilon_{l_1\dots l_d}e^{l_5}\dots e^{l_d}\psi_{\{p\}}{}^{l_1a(k-1),l_2b(l-1)}\phi_{\{q\}}{}^{l_3}{}_{a(k-1)}{}^{,l_4}{}_{b(l-1)},
\end{equation}
where $\epsilon_{l_1\dots l_d}$ is the totally antisymmetric tensor.
The scalar  product
$\langle\dots|\dots\rangle$ is nonzero provided that  the $p$-form
$\psi_{\{p\}}^{a(k),b(l)}$ and $q$-form $\phi_{\{q\}}^{a(k),b(l)}$
have $p+q=4$ and
carry equivalent representations of the Lorentz group
$ \mathbf{Y}(k,l)$, $k\ge l\ge 1$
 (otherwise the r.h.s. of (\ref{1formsgeneral1}) does not make sense).

We impose the following conditions
\begin{gather}
\label{normalization1}
f(k+1,l)=n(k,l)\frac{(d+k-2)k}{k+1}g(k,l),   \quad k\ge 1, l\ge 0,\\
\label{normalization2}
F(k,l+1)=N(k,l)\frac{(d+l-3)(k-l+1)}{(k-l)}G(k,l),   \quad k\ge 1, l\ge 0,
\end{gather}
where the coefficients $n(k,l)$ and $N(k,l)$ will be determined later on.
For $k\ge 1$ and $l\ge 1$ these conditions follow from
\begin{equation*}
\langle \sigma_-^{1}\psi_{\{p\}}(k+1,l)|\phi_{\{q\}}(k,l)\rangle=n(k,l)
(-1)^p \langle \psi_{\{p\}}(k+1,l)|\sigma_+^{1}\phi_{\{q\}}(k,l)
\rangle,  \quad k\ge 1, l\ge 1,
\end{equation*}
\begin{equation}
\label{conjsigma1}
\langle \sigma_-^{2}\psi_{\{p\}}(k,l+1)|\phi_{\{q\}}(k,l)\rangle=N(k,l)
(-1)^p \langle \psi_{\{p\}}(k,l+1)|\sigma_+^{2}\phi_{\{q\}}(k,l)
\rangle,  \quad k\ge 1, l\ge 1\,,
\end{equation}
while  for $k \ge 1$, $l=0$ they are imposed by hand.
 Note that eq.~(\ref{normalization2}) can be assumed to be satisfied
 at $k=l=0$  as well  because
 $F(0,1)=G(1,1)=0$ by (\ref{oblopred}), (\ref{fgzero}).
However, the condition (\ref{normalization1}) cannot be imposed at $k=l=0$
 to relate $f(1,0)$ and $g(0,0)$, because   it would imply that
 $f(1,0)=0$ that is not necessarily true. Instead, it is convenient to impose
 the condition
 \begin{equation}
\label{f10g00}
f(1,0)=n(0,0)g(0,0).
 \end{equation}

 Let us note that Eqs. (\ref{normalization1}), (\ref{normalization2}) give
 \begin{equation}
 \label{Nn1}
  \frac{g(k,l+1)f(k+1,l)G(k,l)F(k+1,l+1)}{g(k,l)f(k+1,l+1)G(k+1,l)F(k,l+1)}=
 \frac{N(k+1,l)n(k,l)}{N(k,l)n(k,l+1)}\frac{(k-l)(k-l+2)}{(k-l+1)^2}.
 \end{equation}
 On the other hand, from (\ref{3.11}) it follows that
  \begin{equation}
  \label{Nn2}
   \frac{g(k,l+1)f(k+1,l)G(k,l)F(k+1,l+1)}{g(k,l)f(k+1,l+1)G(k+1,l)F(k,l+1)}=
 \frac{(k-l)(k-l+2)}{(k-l+1)^2}.
 \end{equation}
 So, the conditions (\ref{conjsigma1}) require
 \begin{equation}
 \label{Nn3}
 \frac{N(k+1,l)n(k,l)}{N(k,l)n(k,l+1)}=1.
 \end{equation}
 Once (\ref{Nn3}) is true, Eq. (\ref{Nn1}) is equivalent to the equation (\ref{Nn2}) which is invariant under the transformations (\ref{3.16}). This explains how it is possible to achieve the two conditions
 (\ref{normalization1}), (\ref{normalization2}) with the help of a single
  function $\beta(k,l)$.
 After imposing (\ref{normalization1}), (\ref{normalization2}), the ambiguity due to
  rescaling (\ref{3.15}) is fixed up to an overall rescaling with $\beta(k,l)=const$.

\subsection{Formal solution}
Let us introduce the following combinations of the coefficients
\begin{equation}
\label{definitionhsm}
h(k,l)=f(k+1,l)g(k,l)\,, \qquad
H(k,l)=F(k,l+1)G(k,l),
\end{equation}
that remain invariant under the redefinitions (\ref{3.15}) and satisfy the following
 two equations  as a consequence of Eqs.~(\ref{3.11})
\begin{equation}
\label{3.17}
\begin{split}
\frac{h(k,l-1)}{h(k,l)}= \frac{k-l+1}{k-l+2}\quad \frac{d+k+l-2}{d+k+l-3},\\
\frac{H(k+1,l)}{H(k,l)}=\frac{d+k+l-2}{d+k+l-1}\quad \frac{k-l+1}{k-l+2}.
\end{split}
\end{equation}
 The condition  (\ref{3.10}) gives
\begin{equation}
\label{3.19}
\begin{split}
h(k,l)\frac{d+2k}{d+2k-2}-h(k-1,l)-H(k,l)\frac{d+2l-2}{(k-l)(d+k+l-3)}-\lambda^2=0,\\
h(k,l)\frac{d+2k}{(k-l+2)(d+k+l-3)}+H(k,l)\frac{d+2l-2}{d+2l-4}-H(k,l-1)-\lambda^2=0.
\end{split}
\end{equation}

Given $H(k,l)$ and $h(k,l)$, $\sigma$ is reconstructed  by
(\ref{normalization1}), (\ref{normalization2}), (\ref{definitionhsm})
\begin{equation}
\label{app01}
f(k+1,l)=\sqrt{n(k,l)\frac{(d+k-2)k}{k+1}h(k,l)}, \quad k>0, \quad l\le k,
\end{equation}
\begin{equation}
\label{app02}
g(k,l)={\rm sign}(h(k,l))\sqrt{\frac{k+1}{k(d+k-2)n(k,l)}h(k,l)},\quad k>0, \quad l\le k,
\end{equation}
\begin{equation}
\label{app025}
f(1,0)=\sqrt{n(0,0)h(0,0)}, \quad g(0,0)={\rm sign}(h(0,0))\sqrt{h(0,0)/n(0,0)},
\end{equation}
\begin{equation}
\label{app03}
F(k,l+1)=\sqrt{N(k,l)\frac{(d+l-3)(k-l+1)}{k-l}H(k,l)}, \quad k>l\ge 0,
\end{equation}
\begin{equation}
\label{app04}
G(k,l)={\rm sign}(H(k,l))\sqrt{\frac{k-l}{(d+l-3)(k-l+1)N(k,l)}H(k,l)}, \quad k>l\ge 0.
\end{equation}

%\subsection{General solution}

The formal general solution of the equations (\ref{3.17}) and (\ref{3.19}) is
    \begin{equation}
      \label{3.21}
    \begin{split}
    h(k,l)=-\frac{(d+2l_0-2)(k_0-k)(d+k_0+k-1)}{(d+2k)(k-l+1)(d+k+l-2)}H(k_0,l_0)+\\
     +\frac{(d+2k_0)(k-l_0+1)(d+k+l_0-2)}{(d+2k)(k-l+1)(d+k+l-2)}h(k_0,l_0)+\\
     -\lambda^2\frac{(k_0-k)(d+k_0+k-1)(k-l_0+1)(d+k+l_0-2)}{(d+2k)(k-l+1)(d+k+l-2)},
     \end{split}
     \end{equation}
     \begin{equation}
     \label{3.21.0}
     \begin{split}
     H(k,l)=\frac{(d+2k_0)(l_0-l)(d+l_0+l-3)}{(d+2l-2)(d+k+l-2)(k-l+1)}h(k_0,l_0)+\\
     +\frac{(d+k_0+l-2)(k_0-l+1)(d+2l_0-2)}{(d+k+l-2)(k-l+1)(d+2l-2)}H(k_0,l_0)+\\
     -\lambda^2\frac{(d+k_0+l-2)(k_0-l+1)(d+l_0+l-3)(l_0-l)}{(d+k+l-2)(k-l+1)(d+2l-2)}.
     \end{split}
     \end{equation}
From Eqs. (\ref{3.21}), (\ref{3.21.0}) we observe that $H(k,l)$ and $h(k,l)$ with $k\ge l\ge 0$ are determined by
their values $H(k_0,l_0)$ and $h(k_0,l_0)$ at any point $(k_0,l_0)$ with $k_0\ge l_0\ge 0$.
That this should have happen follows from  formal consistency  of the
``finite difference equations" (\ref{3.17}), (\ref{3.19}) in the integral
variables $k$ and $l$.
Note that denominators in (\ref{3.21}), (\ref{3.21.0}) are  nonzero
for $k\ge l$, which is the condition that the second row of a Young
diagram is not longer than the first one.

An important property of the solution (\ref{3.21}), (\ref{3.21.0}) is expressed by

{\em Lemma.} If $H(p,s)=0$  or  $h(s-1,p)=0$ for some $p$ and $s$
then $H(k,s)=0 \quad \forall \quad k\,$
and  $h(s-1,l)=0 \quad \forall \quad l$
  \begin{equation}
    \label{boundary}
 H(p,s)=0 \quad \text{or} \quad h(s-1,p)=0\,:\quad \Rightarrow \quad
 H(k,s)=h(s-1,l)=0 \quad \forall \quad k, l\,.
   \end{equation}

{\em Proof.}  Suppose that $H(p,s)=0$. Let us use (\ref{3.21}), (\ref{3.21.0})
 to express all $h(k,l)$ and $H(k,l)$
in terms of $H(p,s)$ and $h(p,s)$ setting $k_0=p$ and $l_0=s$.
The first term in (\ref{3.21}) is then proportional to $H(p,s)$
and hence vanishes. The other terms contain a factor of $(k-s+1)$ that is
zero for $k=s-1$. Hence $h(s-1,l)=0$. The second term in (\ref{3.21.0})
is proportional to $H(p,s)$ and hence vanishes.
The other terms contain a factor of $(s-l)$ which implies that $H(k,s)=0$.
That (\ref{boundary}) is true if $h(s-1,p)=0$  is proved  analogously.  $\Box$

The solution (\ref{3.21}), (\ref{3.21.0}) is formal since it does not take
into account (\ref{fgzero}) which demands
\begin{equation}
\label{boundarydiag}
H(n,n)=h(n-1,n)=0.
\end{equation}
This can be treated as a boundary condition for Eqs.~(\ref{3.17})-(\ref{3.19}).
Generally, (\ref{boundarydiag}) is not true for (\ref{3.21}), (\ref{3.21.0}).
It turns out that if we would use the formula (\ref{3.21}), (\ref{3.21.0}) for all $h$ and $H$
except for those given by (\ref{boundarydiag}), the compatibility condition (\ref{sigmasquared})
will not be satisfied only
 in the sector of rectangular Young diagrams
 \begin{equation}
 \label{inco}
 \left(D^2+\sigma^2\right)\mid_{\cal W}\neq 0, \quad
 \left(D^2+\sigma^2\right)\mid_{V/{\cal W}}= 0\,,
 \end{equation}
 where $\cal W$ denotes the set of operators that
 map rectangular Young diagrams to
 rectangular ones. For example,
 $\sigma_-^1(n,n-1)\sigma_-^2(n,n)\in \cal W$ because it  maps
 $\mathbf{Y}(n,n)$ to $\mathbf{Y}(n-1,n-1)$.

%Let us now analyze the solution (\ref{3.21}), (\ref{3.21.0}) in some more detail.

 In the general case, the equations (\ref{eqhom}) contain $W^{a(k),b(l)}$ with all
 $k\ge l$. $H(k,l)$ and $h(k,l)$ are
 defined by the two-parametric solution (\ref{3.21}), (\ref{3.21.0}). In this case the boundary condition
 (\ref{boundarydiag}) is not satisfied. Hence the curvatures (\ref{eqhom})
 are inconsistent for generic coefficients $H(k,l)$ and $h(k,l)$ that should be chosen
 appropriately to describe a spin $s$ field.

\subsection{Scaling gauges}
\label{scg}
The coefficients $h(k,l)$ (\ref{3.21}) and $H(k,l)$ (\ref{3.21.0})
can be chosen to be real. In order the coefficients $f(k,l)$,
$g(k,l)$, $F(k,l)$ and $G(k,l)$ also be real we demand
\begin{equation}
\label{reality} {\rm sign}(n(k,l))={\rm sign}(h(k,l)), \qquad {\rm
sign}(N(k,l))={\rm sign}(H(k,l))
\end{equation}
in (\ref{normalization1}), (\ref{normalization2}).
The signs in (\ref{app01}), (\ref{app03}) are chosen so
that $f(k,l)$ and $F(k,l)$ are positive.

In the practical analysis we will deal with differential forms of
degrees $0$ and $1$ and use the two ways of fixation of the field
rescaling ambiguity.
 The   {\em type-I scaling gauge} is
\begin{equation}
\label{scaling0} n_{(0)}(k,l)=n_{(1)}(k,l)={\rm sign}(h(k,l)), \quad
N_{(0)}(k,l)=N_{(1)}(k,l)={\rm sign}(H(k,l))\,,
\end{equation}
where the subscripts $(0)$ or $(1)$ indicate a degree of the
differential form in question. Although the type-I scaling is most
convenient for the general analysis, it is inapplicable to the
analysis of the flat massless limit where it should be replaced by
the {\em type-II scaling gauge}
 \begin{equation}
\label{newscaling1} N_{(1)}(k,l)={\rm
sign}(H(k,l))\left|\frac{1}{m^2+\lambda^2(s-l-1)(s+l+d-4)}\right|,
\quad n_{(1)}(k,l)={\rm sign}(h(k,l)),
\end{equation}
\begin{equation}
\label{newscaling11} N_{(0)}(k,l)={\rm sign}(H(k,l)), \quad
n_{(0)}(k,l)={\rm
sign}(h(k,l))\left|\frac{1}{m^2+\lambda^2(s-k-2)(s+k+d-3)}\right|\,.
\end{equation}

  It is easy to check that (\ref{newscaling1}) and (\ref{newscaling11}) respect
  the  condition (\ref{Nn3}). Since $N_{(1)}(k,l)$ is $k$-independent
  and $n_{(0)}(k,l)$ is $l$ independent, we introduce  notation
\be
  N_{(1)}(l)=N_{(1)}(k,l)\q n_{(0)}(k)=n_{(0)}(k,l)\,.
  \ee
Note that \be N_{(1)}(l)=-n_{(0)}(l-1). \ee

\subsection{Spin $s$ systems}
It turns out that solutions of (\ref{3.21}), (\ref{3.21.0}),
 that satisfy (\ref{boundary}) with some $s$, correspond to a spin
 $s$ field.

  \begin{figure}[h]
    \centering
    \includegraphics[width=12cm]{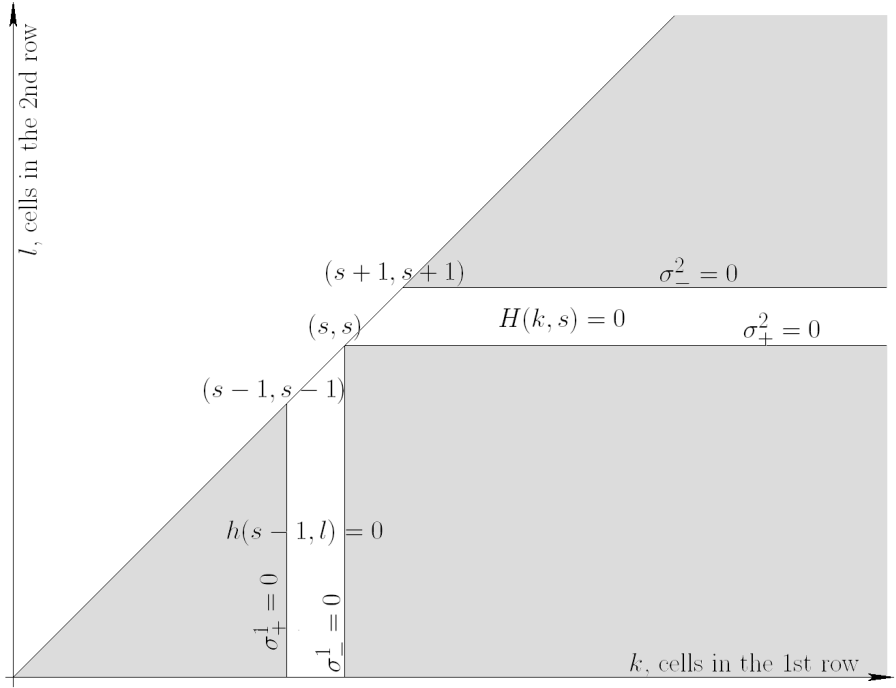}
    \caption{Spin $s$ field, non-critical $m\neq m_t$.
    Fields in distinct shaded regions can form independent unfolded
    systems (\ref{3.2})}
\end{figure}

{}From the condition $h(s-1,l)=0$ and (\ref{definitionhsm}) it follows that
either $f(s,l)=0$ or $g(s-1,l)=0$.
Consider the case where $f(s,l)=g(s-1,l)=0$ for all $l$. It is easy to see that
$f(s,l)=0$ leads to $\sigma_-^1(s,l)=0$, which means that the connections
  $W^{a(u),b(v)}$ with $u\ge s$ do not contribute to the curvatures
  $R^{a(k),b(l)}$ (\ref{3.2}) with $k\le s-1$. Analogously,  $g(s-1,l)=0$ leads
  to $\sigma_+^1(s-1,l)=0$, which means in turn that
  $W^{a(k),b(l)}$ with $k\le s-1$ do not contribute to the curvatures $R^{a(u),b(v)}$
(\ref{3.2})  with $u\ge s$. So, if both $f(s,l)=0$ and $g(s-1,l)=0$ are satisfied the
respective curvatures
  do not mix $W^{a(u),b(v)}$ with $u\ge s$ and $W^{a(k),b(l)}$ with $k\le s-1$.
Hence, in this case $W^{a(k),b(l)}$ with $k\le s-1$ (finite triangle
  region in Fig. 1) and $W^{a(u),b(v)}$ with $u\ge s$ (both infinite triangular
  and infinite rectangular regions in Fig. 1) form independent subsystems.
  Relaxing the conditions $f(s,l)=0$ or $g(s-1,l)=0$
   leads to the mixing of the fields
  $W^{a(k),b(l)}$ with $k\le s-1$ and $W^{a(u),b(v)}$ with $u\ge s$.

   The condition $H(k,s)=0$ can be analyzed analogously. It is easy to see
   that in the case where
  $f(s,l)=g(s-1,l)=F(k,s+1)=G(k,s)=0$ for some $s$ and all $k$ and $l$,
   the shaded regions in Fig. 1 form  independent
  subsystems that are not mixed in the curvatures (\ref{3.2}).

  The subsystem associated to the infinite rectangular region in Fig. 1
  contains $W^{a(k),b(l)}$ with $k\ge s$, $l\le s$.
  In the case of 0-forms ($p=0$), it describes a so-called
  Weyl module of a massive  particle of spin $s$ that encodes
  nontrivial degrees of freedom of the system. The corresponding
  coefficients $H(k,l)$ and $h(k,l)$ are
  \begin{equation}
\label{3.39}
\begin{split}
  h(k,l)=-\frac{\left(m^2+\lambda^2(s-k-2)(s+k+d-3)\vphantom{\frac ab}\right)(k-s+1)(d+s+k-2)}{(d+2k)(k-l+1)(d+k+l-2)}, \\
 H(k,l)=-\frac{\left(m^2+\lambda^2(s-l-1)(s+l+d-4)\vphantom{\frac ab}\right)(s-l)(d+s+l-3)}{(d+2l-2)(k-l+1)(d+k+l-2)},
 \end{split}
\end{equation}
where the free parameter $m$ identifies with the mass as one can see by the
direct analysis of the equations (\ref{3.2}) for 0-forms $C^{a(k),b(l)}$
with $k\ge s$, $l\le s$ analogous to the analysis  in the 1-form sector
 carried out in Section 3.3.2.
Note that Eq.~(\ref{3.39}) provides an  alternative
 parametrization of the general solution (\ref{3.21}), (\ref{3.21.0})
 with $s$ interpreted as an arbitrary (not necessarily integer) parameter.

 \begin{figure}[h]
    \centering
    \includegraphics[width=12cm]{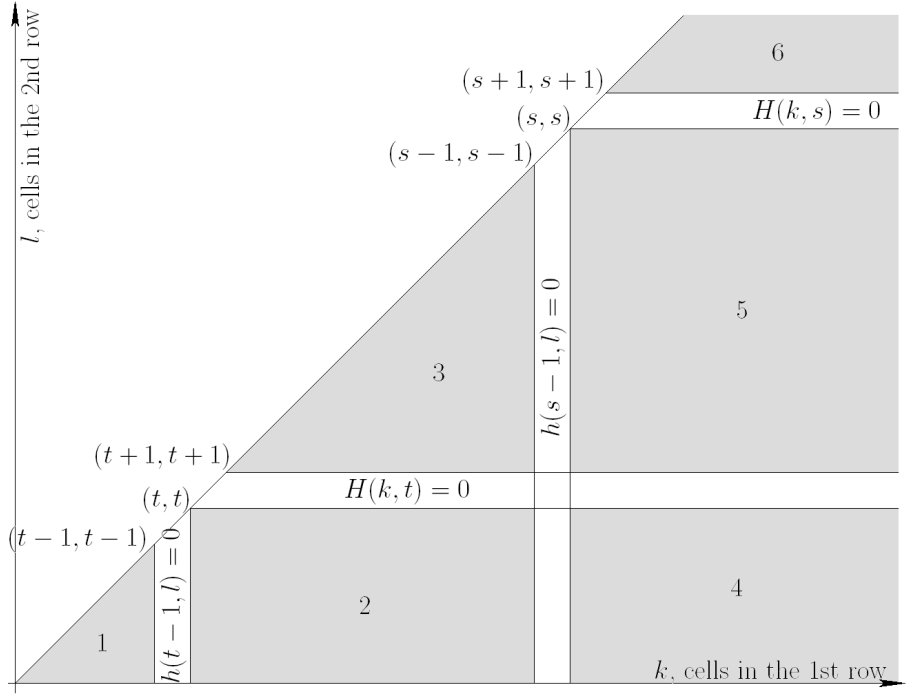}
    \caption{Partially massless field, $m=m_t$. The set of fields can be divided into six groups}
\end{figure}

Both finite ($k\le s-1$) and infinite ($l>s$) triangle regions in Fig. 1 do not respect
(\ref{boundarydiag}). Nevertheless, in Section 3 we will show that
the subsystem associated to the finite triangle plays important role in the description
of massive fields, leading to consistent unfolded equations (\ref{eqhom}) upon
introducing an appropriate mixture between $0$- and $1$-forms.

 It is easy to see that for special values $m=m_t$ ($t$ is integer) in (\ref{3.39})
 \begin{equation}
 \label{partiallymassless}
 m_t^2=-\lambda^2 (s-t-1)(d+s+t-4), \quad t=0,1,\dots,s-1
 \end{equation}
the following property takes place
\begin{equation}
\label{pmbc}
H(k,t)=h(t-1,l)=0.
\end{equation}
Analogously to the condition (\ref{boundary}),
Eq.~(\ref{pmbc}) subdivides the system into further subsystems as
 shown in Fig. 2. Since the region 4 does not contain
fields valued in $\mathbf{Y}(n,n)$  the condition (\ref{boundarydiag}) here
is inessential. The regions 2 and 5 both respect (\ref{boundarydiag})
as a consequence of (\ref{boundary}) and (\ref{partiallymassless}). To summarize, the regions
2, 4 and 5 respect (\ref{sigmasquared})  leading to consistent curvatures,
while in the regions
1, 3 and 6 the condition (\ref{boundarydiag}) is not respected.

The critical values of mass (\ref{partiallymassless})
correspond to partially massless \cite{Deser, ZinovievMetric, SkvortsovPM} and massless
 \cite{Vasiliev1987, LopatinVasiliev} fields for $t<s-1$ and
$t=s-1$, respectively. Namely, the subsystem associated to the region 2  of Fig. 2
was used to describe partially massless fields in terms of 1-forms in \cite{SkvortsovPM}.
The region 5 describes the corresponding Weyl module.

Note that letting $t,s \in \mathbb R$ in Eqs. (\ref{3.39}), (\ref{partiallymassless})
provides an alternative parametrization of the general solution
(\ref{3.21}), (\ref{3.21.0}).

Also, from (\ref{3.39}) it is easy to see that if
$m^2$ is given by (\ref{partiallymassless}) with integer
$t$, $t>s$ and
$\lambda^2\ne 0$ then the Weyl module of such a
field satisfies (\ref{pmbc}). In this case the Weyl module (the infinite
rectangular region in Fig. 1) admits a submodule.
 It consists of 0-forms $C^{a(k),b(l)}$ with $s\le k\le t-1$.
This phenomenon was observed for the case of massive spin 0 field in \cite{Shayns0}.
Such degeneracies usually indicate the coexistence of different dual descriptions
of the same field theoretical system, and hence may be interesting from various
perspectives.

The following comment is now in order. It is well-known (see e.g.
\cite{SolvayWorkshop} and references therein)
that any consistent unfolded equation for a set of
forms of a definite degree implies that this set spans a
module of the global symmetry of the system, which is Poincar\'{e}
or $(A)dS$ symmetry in the case of most interest in this paper. That the compatibility
conditions are not satisfied only at the boundary of the area of
definition of Young tableaux suggests the following interpretation.
Gauge fields of massless and partially massless fields are described
by finite dimensional tensor $o(d-1,2)$-modules $T$ \cite{LopatinVasiliev, SkvortsovPM}. In this case, the parameter
of mass is discrete, being associated to the length of the second row of
the respective Young tableaux \cite{SkvortsovPM}. These finite dimensional
modules can be understood as particular members of the set of
modules $V_m$ parametrized by the continuous parameter of mass $m$.
More precisely, for the specific values $m_i$ of $m$ associated
to massless and
partially massless fields, $V_{m_i}$ acquire infinite dimensional
submodules $U_{m_i}$ such that $V_{m_i}/U_{m_i}=T_i$ are the finite dimensional
modules used for the description of (partially) massless fields.
That the unfolded equations are compatible almost everywhere except for
the boundary of the region associated to the Young tableaux (i.e.
finite dimensional representations of the Lorentz group) confirms this
picture. Most likely, the obstruction to the continuation of the construction at the
boundary of the Young tableaux region, shown in this section for
generic values of mass, signals that the
corresponding infinite dimensional module $V_m$ does not decompose
into a sum of finite dimensional modules of the Lorentz algebra
as is also anticipated for  the submodules $U_{m_i}$. In other words,
although the module $V_m$ exists, it may admit no interpretation in
the standard field-theoretical terms of finite-component Lorentz fields.

\section{Unfolded equations for massive fields of any spin}
\subsection{Field pattern}
Unfolded equations (\ref{3.2})
formulated in terms of 0-forms $C^{a(k),b(l)}$ with $k\ge s$ and $l\le s$
 (infinite rectangular region in Fig. 1) are consistent. They describe the
 Weyl module of a massive HS field, that operates with gauge invariant
 combinations of gauge fields.
To figure out how to introduce gauge potentials, it is
useful to use the observation of \cite{ZinovievMetric, ZinovievFrame}
that the Lagrangian of a spin
 $s$ massive field can be represented as a sum of the Lagrangians
 for the set of spin $s'$, $0 \le s'\le s$ massless fields
 supplemented with
 certain mass-dependent lower-derivative terms that mix massless fields
 of different spins. Hence, it is natural to expect that the unfolded formulation
of a massive spin $s$ particle can be given in terms of fields
of the unfolded formulation for the set of massless fields of spins $0\le s'\le s$.

 \begin{figure}[h]
    \centering
    \includegraphics[width=12cm]{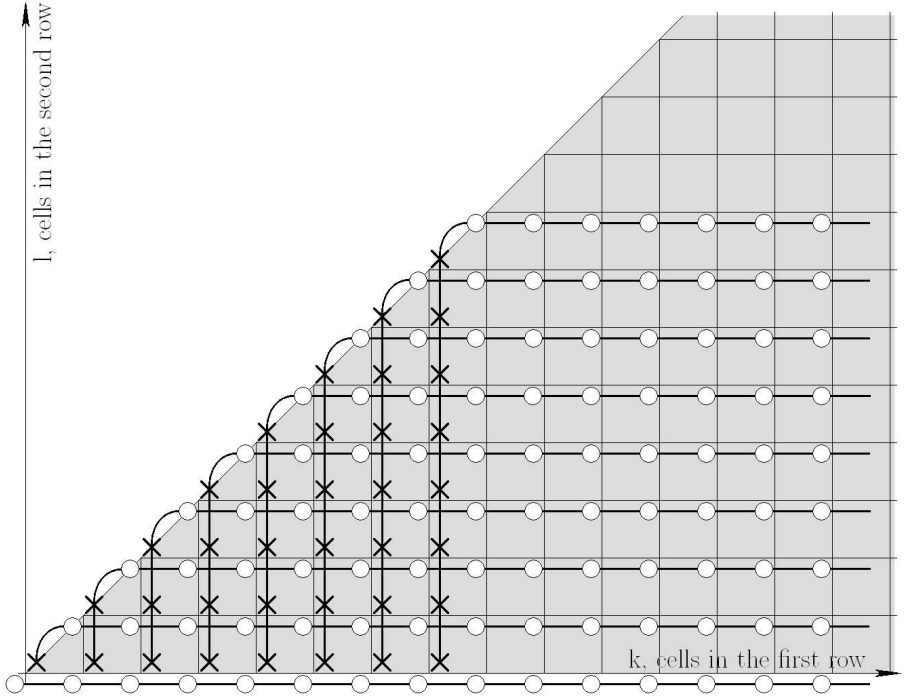}
    \caption{On this plot the set of fields used for description of a
    massive spin $s$ field is shown for $s=8$.``$\times$" denote 1-forms and
    ``$\circ$" denote 0-forms. The fields associated to each massless field are
    connected by a solid line. The set of fields of a massive spin $s$ model is
    a combination of the sets of fields of the massless unfolded systems of spins
    $s^\prime= 0,\ldots s$.}
\end{figure}

 As shown in \cite{LopatinVasiliev, Vasiliev2001},  a spin $s'$ symmetric massless
 field is described by the set of 1-form connections valued in traceless tensors
 of $o(d-1,1)$ that have symmetries of  two-row Young tableaux with
 $ s'-1$ cells in the first row and  $0\le l\le s'-1$ cells in the second row
 \begin{equation*}
 \text{1-forms:} \qquad \RectBRow{5}{3}{$s'-1$}{$l$},\qquad 0\le l\le s'-1
 \end{equation*}
  and Weyl 0-forms  valued in traceless tensors of $o(d-1,1)$ possessing symmetries of
 two-row Young tableaux with  $k \geq s'$ cells in the first row and
 $s'$ cells in the second row
 \begin{equation*}
 \text{0-forms:} \qquad \RectBRow{5}{3}{$k$}{$s'$},\qquad k\ge s'.
 \end{equation*}

 Thus the full set of fields appropriate for
 the description of a spin $s$ massive field should
 contain 1-form connections valued in $\mathbf{Y}(k,l)$ with $k\le s-1$,
 $l\le k$ and 0-forms valued in $\mathbf{Y}(k,l)$ with
 $l\le s$, $k\ge l$. We observe that the resulting set of 1-forms
precisely corresponds to the triangular region of Fig. 3 while the set of
 0-forms corresponds to the unification of the infinite rectangular region
 with the triangular region of Fig. 3. In this section it will be shown how
 the  compatibility problems in the triangular region
 encountered in the case  of forms of definite degree are
 resolved by  a mixture between $1$-forms and $0$-forms of this region.

  Note that the gluing  the Weyl module to the
combined set of 1- and 0-forms in the triangle is not necessary
for formal  consistency but is needed to relate the 1-forms
to the propagating degrees of freedom described by the 0-form
Weyl module associated to the infinite rectangular region.

\subsection{Unfolded equations via mixing of forms of different degrees}
\label{1and0forms}

General  linear unfolded equations (\ref{2.1}), (\ref{2.2}) formulated
 in terms of $(p+1)$-forms $W^{a(k),b(l)}$ and $p$-forms $C^{a(k),b(l)}$,
that possess symmetries of two-row  Young tableaux,
 have the  structure
 \begin{equation}
 \label{4.1}
 \begin{split}
 0=R_{\{p+2\}}^{a(k),b(l)}=dW^{a(k),b(l)}+G^{a(k),b(l)}_{(p+1)}(W,C)=dW+ \sigma_{(p+1)}W+\kappa C,\\
 0=R_{\{p+1\}}^{a(k),b(l)}=dC^{a(k),b(l)}+G^{a(k),b(l)}_{(p)}(W,C)=dC+\sigma_{(p)}C+\alpha W.
 \end{split}
 \end{equation}
A label in parentheses refers to the differential form degree of a curvature.
Since $\alpha$ carries differential form degree 0,
 it is independent of the vielbein and does not
 change a tensor type, \ie
 \be
 \ga \mathbf{Y}(k,l)= \ga(k,l)\mathbf{Y}(k,l),
 \ee
 where $\ga(k,l)$ are some coefficients.
 (Here we use the shorthand notation $\mathbf{Y}(k,l)$ for a tensor
 described by the Young tableau with two rows of lengths $k\geq l$.)

The operators $\kappa$ are
 built from two vielbeins, carrying the differential form degree two and contain
 several terms
 \begin{equation}
 \label{kappacomponenets1}
 \kappa=\kappa_0+\kappa_{--}^{12}+\kappa_{++}^{12}+\kappa_{+-}^{12}+\kappa_{-+}^{12}\,,
 \end{equation}
 \begin{equation*}
 \kappa_0: \mathbf{Y}(k,l)\rightarrow \mathbf{Y}(k,l), \quad
\kappa_{--}^{12}: \mathbf{Y}(k,l)\rightarrow \mathbf{Y}(k-1,l-1),
 \end{equation*}
 \begin{equation}
 \label{kappacomponenets}
 \kappa_{++}^{12}: \mathbf{Y}(k,l)\rightarrow \mathbf{Y}(k+1,l+1),  \quad
\kappa_{-+}^{12}: \mathbf{Y}(k,l)\rightarrow \mathbf{Y}(k-1,l+1),
 \end{equation}
 \begin{equation*}
 \kappa_{+-}^{12}: \mathbf{Y}(k,l)\rightarrow \mathbf{Y}(k+1,l-1)\,.
 \end{equation*}
 Each of these operators except for $\kappa_0$ is determined by the Young
symmetry properties up to a $(k,l)$-dependent coefficient. $\kappa_0$ contains
two linearly independent terms, each defined up to a $(k,l)$-dependent coefficient.

 The operators $\sigma_{(p+1)}$ and $\sigma_{(p)}$  act on $(p+1)$-forms and
$p$-forms, respectively and add one  differential form degree. They still have the form (\ref{3.5})-(\ref{3.5.3})
  but the compatibility conditions (\ref{2.3}) are modified because of presence of $\kappa$ and $\alpha$.
  Hence the respective
  coefficients $F_{(p+1)}(k,l)$ ($F_{(p)}(k,l)$), $G_{(p+1)}(k,l)$
  ($G_{(p)}(k,l)$), $f_{(p+1)}(k,l)$ ($f_{(p)}(k,l)$) and $g_{(p+1)}(k,l)$
  ($g_{(p)}(k,l)$) may differ from those given in (\ref{definitionhsm})
   and (\ref{3.21}), (\ref{3.21.0}).

   In the new framework, the compatibility conditions imply
   \begin{equation}
\label{4.3}
   \alpha\sigma_{(p+1)}=\sigma_{(p)}\alpha,
   \end{equation}
\begin{equation}
\label{4.4}
\alpha\kappa=D^2+(\sigma_{(p)})^2,
\end{equation}
\begin{equation}
\label{4.5}
\kappa\alpha=D^2+(\sigma_{(p+1)})^2,
\end{equation}
\begin{equation}
\label{4.6}
\sigma_{(p+1)}\kappa=\kappa\sigma_{(p)}.
\end{equation}
Eqs. (\ref{4.4}), (\ref{4.5}) express $\kappa$ in terms of $\sigma$.

{}From (\ref{4.3}) it follows  that
\begin{equation}
\label{hfordifferentforms1}
h_{(p+1)}(k,l)=h_{(p)}(k,l)\qquad \text{if} \qquad  \alpha(k,l) \alpha(k+1,l)\ne 0,
\end{equation}
\begin{equation}
\label{hfordifferentforms2}
H_{(p+1)}(k,l)=H_{(p)}(k,l)\qquad \text{if} \qquad  \alpha(k,l) \alpha(k,l+1)\ne 0.
\end{equation}
Indeed, let us for example derive (\ref{hfordifferentforms1}).
The operators $\sigma_-^1(k+1,l)$ and $\sigma_+^1(k,l)$ for $p$- and $(p+1)$-forms are related by the
conditions (\ref{4.3})
\begin{equation*}
\alpha(k,l)\sigma_{(p+1)}{}_-^1(k+1,l)=\sigma_{(p)}{}_-^1(k+1,l)\alpha(k+1,l),
\end{equation*}
\begin{equation*}
\alpha(k+1,l)\sigma_{(p+1)}{}_+^1(k,l)=\sigma_{(p)}{}_+^1(k,l)\alpha(k,l).
\end{equation*}
This implies
\begin{equation*}
\alpha(k,l)f_{(p+1)}(k+1,l)=f_{(p)}(k+1,l)\alpha(k+1,l),
\end{equation*}
\begin{equation}
\label{hfordifferentforms3}
\alpha(k+1,l)g_{(p+1)}(k,l)=g_{(p)}(k,l)\alpha(k,l).
\end{equation}
Multiplying these relations and using (\ref{definitionhsm}) we obtain
(\ref{hfordifferentforms1}).

The gauge transformations that follow from (\ref{2.6}) are
 \begin{equation}
\label{gaugewc}
 \delta W=d\xi+\sigma_{(p+1)}\xi, \qquad
\delta C=-\alpha\xi.
\end{equation}

In addition to the independent scaling redefinitions (\ref{3.15}) for $p$- and $(p+1)$-forms,
 the equations (\ref{4.1}) possess the ambiguity in the field redefinitions
 \begin{equation}
\label{4.7}
  W=\tilde{W}+\tilde{\sigma}C
 \end{equation}
that leads to following redefinition of the operators
 \begin{equation}
\label{ambiguity}
 \sigma_{(p)}\rightarrow\sigma_{(p)}+\alpha\tilde{\sigma}, \quad
  \sigma_{(p+1)}\rightarrow\sigma_{(p+1)}+\tilde{\sigma}\alpha, \quad \kappa\rightarrow \kappa+\tilde\sigma\sigma_{(p)}+\sigma_{(p+1)}\tilde\sigma+\tilde\sigma\alpha\tilde\sigma.
 \end{equation}

It is not hard to see that to
 single out a solution of the compatibility conditions that describes a
spin $s$ particle in the case of $p=0$, the following condition, that plays
a role  analogous to that of (\ref{boundary}), should be imposed
 \begin{equation}
 \label{specificmassive}
 \alpha(k,l)=0 \quad \text{for} \quad k\ge s, \quad \alpha(k,l)\neq 0 \quad \text{for} \quad k\le s-1\,.
 \end{equation}
Then from (\ref{4.3}) it follows that
 \begin{equation}
 \label{granusl}
 \sigma_{(0)}{}_+^1(s-1,l)=\sigma_{(1)}{}_-^1(s,l)=0\,.
 \end{equation}

Indeed, the condition (\ref{specificmassive}) and the second
of the conditions (\ref{granusl}) imply that
 $1$-forms $W^{a(u),b(v)}$ with $u\ge s$ do not contribute to the
 equations for all 0-forms and 1-forms $W^{a(k),b(l)}$ with $k\le s-1$. Hence,
  (\ref{specificmassive}) guarantees that
 the 1-forms in the  infinite triangular and rectangular regions of Fig. 3 decouple from
 all other fields and can be dropped from the system.
Since the curvatures for 0-forms valued in $\mathbf{Y}(u,v)$ with
 $u\ge s$ do not contain 1-forms, the analysis of Section 2 is applicable to this case.
The first of the relations (\ref{granusl}) implies that (\ref{boundary}) is satisfied.
This means that the 0-forms in the
 infinite triangular region of Fig. 1 also decouple and can be dropped from the system.
 As a result we are left with the set of fields of Fig. 3.

 Let us now consider the 1- and 0-forms in the triangular region of Fig. 3.
  These should be mixed
  by the equations (\ref{4.1}) to resolve the compatibility
  problem of the unmixed case of Section 2. Together with the Weyl module
  these fields should lead to an unfolded system that
  consistently describes massive HS particles. To ensure that the Weyl module
  is glued to the sector of mixed 0- and 1-forms we demand
  \begin{equation}
  \label{weylgluing}
  \sigma_{(0)}{}_-^1(s,l)\ne 0.
  \end{equation}
  The coefficients $f_{(0)}(s,l)$ will be specified later.

Note that (\ref{ambiguity}) implies that any $\sigma_{(0)}$ (or $\sigma_{(1)}$
related to $\sigma_{(0)}$ by (\ref{4.3})) can be obtained
by some field redefinition (\ref{4.7}) in the sector where 1- and 0-forms are
mixed.
Our goal is to fix this freedom in such a way that, for particular
values of $m$ and $\lambda$ that correspond to massless and partially massless
fields, the unfolded system decompose into appropriate subsystems.
 The solution
 \begin{equation}
 \label{Hh01}
 H_{(0)}(k,l)=H_{(1)}(k,l)=H(k,l), \qquad h_{(0)}(k,l)=h_{(1)}(k,l)=h(k,l)
 \end{equation}
 with $H$ and $h$ given by (\ref{3.39})
satisfies this requirement.
The compatibility problems encountered
 in Section \ref{samedegree} are
resolved by the nontrivial mixture of 1- and 0-forms.
Recall that $H$ and $h$ (\ref{3.39}) solve (\ref{sigmasquared}) on its
  restriction to $\bar{\cal{W}}$, while the restriction of
  $D^2+\sigma^2$ to $\cal{W}$ is nonzero (\ref{inco}). Taking into account
  (\ref{4.4}), we see that
  with this choice of $H$ and $h$ we should set $\kappa=0$ if
  $\kappa\in\bar{\cal{W}}$, while $\kappa|_{\cal{W}}$ should be non-zero.
%This allows to carry out the $\sigma_-$ cohomology analysis as in the Section \ref{gaugeaway}.

To summarize, the unfolded equations that describe the Weyl
module in terms of 0-forms $C^{a(k),b(l)}$ with $k\ge s$, $l\le s$  are
\begin{equation}
\label{finaleq4}
R_{\{1\}}=DC+\sigma_{(0)}C=0 \quad \text{for} \quad R_{\{1\}}^{a(k),b(l)}\quad \text{with} \quad k\ge s,\quad l\le s.
\end{equation}
These fields belong to the infinite rectangular region of Fig. 3.
 The finite triangle region contains 1-forms
$W^{a(k),b(l)}$ and 0-forms $C^{a(k),b(l)}$ with $k\le s-1$ that satisfy
the equations
\begin{equation}
\label{finaleq1}
 R_{\{2\}}=DW+ \sigma_{(1)}W=0, \quad \text{for}\quad R_{\{2\}}^{a(k),b(l)}
 \quad \text{with}\quad l<k\le s-1\,,
 \end{equation}
 \begin{equation}
\label{finaleq2}
 R_{\{2\}}=DW+ \sigma_{(1)}W+\kappa C=0, \quad \text{for} \quad R_{\{2\}}^{a(n),b(n)}
 \quad \text{with} \quad n\le s-1\,,
 \end{equation}
\begin{equation}
\label{finaleq3}
R_{\{1\}}=DC+\sigma_{(0)}C+\alpha W=0 \quad \text{for} \quad
R_{\{1\}}^{a(k),b(l)}\quad \text{with} \quad k\le s-1\,.
\end{equation}
Both $\sigma_{(0)}$ and $\sigma_{(1)}$ are still
defined by (\ref{definitionhsm}) and (\ref{3.39}) up to the
rescaling ambiguity (\ref{3.16}) (independently for 0- and 1-forms).
Explicit formulae for the coefficients, that satisfy the normalization conditions
(\ref{normalization1}), (\ref{normalization2}) with
 arbitrary $N$ and $n$ (\ref{conjsigma1}) that respect (\ref{Nn3}), are
given in Appendix B.

For the type-I scaling gauge (\ref{scaling0}),  $f_{(0)}(s,l)$ is determined
by (\ref{3.11}) up to an overall factor which can be fixed in such a way
that
   \begin{equation}
  \label{weylgluing1}
  f_{(0)}(s,l)=\sqrt{\frac{1}{(s-l)(d+s+l-3)}}\,.
  \end{equation}
Since in the type-I scaling gauge $\sigma_{(0)}=\sigma_{(1)}$ by (\ref{scaling0}), (\ref{Hh01}),
 we obtain from (\ref{4.3}) $\alpha(k,l)=\alpha(0,0)$. By the leftover scaling symmetry we can set
 \begin{equation}
 \label{alphaunity0}
 \alpha(k,l)=1.
\end{equation}

For the type-II scaling gauge (\ref{newscaling1}), (\ref{newscaling11}),
$f_{(0)}(s,l)$ is still
defined by (\ref{weylgluing1}) while the expression for $\alpha(k,l)$ in
terms of $\alpha(0,0)$ acquires the form
\begin{equation*}
\alpha(k,l)=\alpha(0,0)\sqrt{\left|\frac{N_{(1)}(0)}{\prod_{i=l}^k{N_{(1)}(i)}}\right|}.
\end{equation*}
The leftover scaling symmetry is used to fix $\alpha(0,0)\sqrt{|N_{(1)}(0)|}=1$,
so that the
final expression for $\alpha$ is
\begin{equation}
 \label{alphaunity}
\alpha(k,l)=\frac{1}{\sqrt{\left|\prod_{i=l}^k{N_{(1)}(i)}\right|}}.
\end{equation}

The explicit form of $\kappa$ (\ref{kappacomponenets}) is
\begin{equation*}
\kappa_{--}^{12}(n,n)=\frac{1}{\alpha(n,n)}{\sigma_{(1)}{}_-^1(n,n-1)\sigma_{(1)}{}_-^2(n,n)},
\end{equation*}
\begin{equation}
\label{nonzerokappa}
\kappa_{++}^{12}(n,n)=\frac{1}{\alpha(n,n)}{\sigma_{(1)}{}_+^2(n+1,n)\sigma_{(1)}{}_+^1(n,n)},
\end{equation}
\begin{equation*}
\kappa_{0}(n,n)=\frac{1}{\alpha(n,n)}\big(D^2+\sigma_{(1)}{}_-^1(n+1,n)\sigma_{(1)}{}_+^1(n,n)+\sigma_{(1)}{}_+^2(n,n-1)\sigma_{(1)}{}_-^2(n,n)\big)\notag,
\end{equation*}
where $D^2$ is given in Eq. (\ref{dsq}). Other components of $\kappa$ in
 (\ref{kappacomponenets1}),
(\ref{kappacomponenets}) are zero. Note that the condition
$\kappa_{--}^{12}(s,s)\ne 0$ is required by
(\ref{weylgluing}) rather than by the compatibility conditions.

The gauge transformations for 0- and 1-forms in the
 triangular region of Fig. 3 are given by (\ref{gaugewc})
with $p=0$, while the Weyl 0-forms in the rectangular region are gauge invariant.

\subsection{Dynamical fields and Singh-Hagen system}
\label{gaugeaway}

To show that the constructed unfolded system with $m\ne m_t$
(the case of partially massless fields is special and
will be considered in Section \ref{massless}) indeed describes a
massive HS field, it is convenient to use the $\sigma_-$
cohomology analysis \cite{Shayns0,SolvayWorkshop}
with the $\sigma_-$ complex defined precisely as in the massless
case $m=0$ for the set of massless fields of spins $0,1,\ldots s$.
This implies that
the $\sigma_-$ cohomological  grade for $0(1)$-forms identifies with the
 length of the first (second) row of their Young diagrams.
The 1-form fields of the lowest grade
described by one-row diagrams are called  frame-like fields.

Since so defined $\sigma_-$ is independent of $m$,
such analysis treats $m^2$ as a deformation parameter. Hence it treats
constraints like $L\phi_1 +m^2 \phi_2=0$ as dynamical equation
as they are in the massless case of $m^2=0$,
rather than constraints that express $\phi_2$ via $\phi_1$. Such
constraints should therefore be taken into account
after performing the $\sigma_-$ cohomology analysis to show that
the dynamical field $\phi_{a(s)}$ satisfies the equations (\ref{metriceq})
while the other fields are either gauged away or
expressed in terms of dynamical field by the equations of motion.

The $\sigma_-$ cohomology analysis for a massless field  gives the following results
(see e.g.
\cite{SolvayWorkshop}).
Dynamical fields for every massless field of spin $k+1$ contained
in the frame-like field $W^{a(k)}$ consist of two irreducible
symmetric tensors
\begin{equation}
\label{1formdecompsymm}
W^{a(k)}\rightarrow\RectARow{5}{$k+1$}\oplus\RectARow{5}{$k-1$}\,, \quad 0\le k \le s-1.
\end{equation}
A massless spin zero field is described by the scalar 0-form $C$.
This set of fields coincides with that proposed by
Zinoviev in \cite{ZinovievMetric}.

The leftover gauge symmetry is:
\begin{equation}
\label{gaugeleftover}
\delta (W(k,0))^{a(k)}=D(\xi(k,0))^{a(k)}+(\sigma_{(1)}{}_-^1\xi(k+1,0))^{a(k)}+(\sigma_{(1)}{}_+^1\xi(k-1,0))^{a(k)}.
\end{equation}
For generic values of $m$, the first component in (\ref{1formdecompsymm}) of each frame-like field
except for $W^{a(s-1)}$ can be gauge fixed to zero by the
gauge parameter $\xi^{a(k+1)}$  using the second term on the r.h.s.
of (\ref{gaugeleftover}).
The remaining gauge parameter $\xi$ is used to gauge away the 0-form $C$. After the
 gauge symmetry is completely fixed, the remaining non-zero field projections are
\begin{equation}
\label{shfields01}
W^{a(k)}\rightarrow\RectARow{5}{$k-1$}\,,\quad 0<k<s-1,
\end{equation}
\begin{equation}
\label{shfields}
W^{a(s-1)}\rightarrow\RectARow{5}{$s-2$} \oplus \RectARow{5}{$s$}\,.
\end{equation}
This is the  set of fields of Singh and Hagen \cite{SinghHagen}.
All other  fields involved in the
unfolded formulation either are  pure gauge or are  expressed via
derivatives of the Singh-Hagen fields.

Note that for special values of the mass $m$ the second term in the
transformation law (\ref{gaugeleftover}) may degenerate not allowing to
gauge fix to zero the highest component field in
(\ref{1formdecompsymm}) for one or another $k$. This
corresponds to the cases
of massless and partially massless fields considered in more detail in
Section \ref{massless} while in this section we assume that
this does not happen.

\subsubsection{Vanishing of lower-spin supplementary fields}
\label{vanishigls}
Let us show that, in the chosen gauge, all fields (\ref{shfields01}), (\ref{shfields})
except for traceless symmetric projection of $W^{a(s-1)}$
vanish by virtue of the equations of motion. We shall use two equations for
each frame-like field. The first one expresses the Lorentz-like auxiliary field via
the  frame-like field. The second one imposes a non-trivial condition on the frame-like
 field. For the  lower-spin supplementary fields (i.e.
those, associated to the massless spin $s'$ fields with $s'<s$) this
 condition equates them to zero. For the dynamical field
 associated to the massless spin $s$ field it imposes the dynamical equation.

 The proof will be given by induction. Namely,
 assuming that $W^a=0$ and $W^{a,b}=0$  we prove that
 \begin{equation}
\label{sh7}
W^{a(l)}=0, \quad W^{a(l),b}=0 \quad \forall l<k\quad \Longrightarrow \quad W^{a(k)}=0, \quad W^{a(k),b}=0.
\end{equation}

Let us first consider the special cases of $s'=0,\,1,\,2$.
Using the gauge conditions
\be
\label{gc}
C(0,0)=0\q W(0,0)=0\,,
\ee
the lowest equation in the spin zero
sector
$$0=DC(0,0)+\sigma_{(0)}{}_-^1C(1,0)+ \alpha(0,0)W(0,0)$$
leads to
\begin{equation}
\label{sh1}
C^a=0.
\end{equation}
The only non-zero projection of $W^a$ is its trace $W_{\rho|}{}^{\rho}$ since,
according to (\ref{shfields01}),
all other its components have been gauged fixed to zero.
From the trace of the equation
\begin{equation}
\label{sh2}
0=D(C(1,0))^a+(\sigma_{(0)}{}_-^1C(2,0))^{a}+(\sigma_{(0)}{}_+^1C(0,0))^a+(\sigma_{(0)}{}_-^2C(1,1))^a+\alpha(1,0)(W(1,0))^a
\end{equation}
we obtain  using (\ref{gc}) that $W_{\rho|}{}^{\rho}=0$ and hence
\begin{equation}
\label{sh3}
W^a=0.
\end{equation}
Also it follows from (\ref{sh2}) that
\begin{equation}
\label{sh4}
C^{a,b}=0.
\end{equation}

To show that $W^{a,b}=0$, we observe that $W_{\mu|}{}^{a,b}$ has three
irreducible components
\begin{equation}
\label{wab1}
{ \atop W^{a,b}}{ \atop\rightarrow} { \atop\YoungBA} { \atop\oplus}{ \atop\YoungAAA}
{ \atop\oplus}{ \atop\YoungA}{ \atop .}
\end{equation}
Projecting the equation
\begin{equation}
\label{sh5}
0=D(W(1,0))^a+(\sigma_{(1)}{}_-^1W(2,0))^{a}+(\sigma_{(1)}{}_+^1W(1,0))^a+(\sigma_{(1)}{}_-^2W(1,1))^{a}
\end{equation}
to the symmetry types of the first two terms in (\ref{wab1})
we obtain that the corresponding projections of $W^{a,b}$ are zero.
Taking into account (\ref{sh4}), the trace of the equation
\begin{equation*}
0=D(C(1,1))^{a,b}+(\sigma_{(0)}{}_-^1C(2,1))^{a,b}+(\sigma_{(0)}{}_+^2C(1,0))^{a,b}+\alpha(1,1)(W(1,1))^{a,b}
\end{equation*}
 implies that the third component of $W^{a,b}$ in (\ref{wab1})
is also zero. Hence,
\begin{equation}
\label{sh6}
W^{a,b}=0.
\end{equation}

The equations (\ref{sh3}), (\ref{sh6}) provide
the first step of induction. Let us now prove (\ref{sh7}).

The projection of the equation
\begin{equation}
\label{sh8}
\begin{split}
0=D(W(k-1,0))^{a(k-1)}+(\sigma_{(1)}{}_-^1W(k,0))^{a(k-1)}+\qquad \\
 +(\sigma_{(1)}{}_+^1W(k-2,0))^{a(k-1)}+(\sigma_{(1)}{}_-^2W(k-1,1))^{a(k-1)}
\end{split}
\end{equation}
to the symmetry type of $\RectARow{5}{$k-1$}$ implies that the second component of $W^{a(k)}$
 in (\ref{1formdecompsymm}) is zero. Since the first component has been gauge fixed to zero,
 we obtain
\begin{equation}
\label{sh9}
W^{a(k)}=0.
\end{equation}

 The situation with $W^{a(k),b}$ is a bit more involved.
 It has the following irreducible components
 $${  \atop W^{a(k),b}}{  \atop\rightarrow}{  \atop\RectBYoung{5}{$k$}{\YoungB}}{  \atop\oplus}
 {  \atop\RectBRow{5}{1}{$k-1$}{$ $}}{ \atop\oplus}
{  \atop\RectBRow{5}{1}{$k+1$}{$ $}}{  \atop\oplus}{  \atop\RectCRow{5}{1}{1}{$k$}{$$}{$$}}{  \atop\oplus}{  \atop\RectARow{5}{$k$}}{ \atop .}$$
The component ${  \atop\RectBYoung{5}{$k$}{\YoungB}}$
is pure gauge and can be gauge fixed to zero as mentioned
in the beginning of Section \ref{gaugeaway}.
 The projections of the equation
\begin{equation}
\label{sh10}
0=D(W(k,0))^{a(k)}+(\sigma_{(1)}{}_-^1W(k+1,0))^{a(k)}+(\sigma_{(1)}{}_+^1W(k-1,0))^{a(k)}+(\sigma_{(1)}{}_-^2W(k,1))^{a(k)}
\end{equation}
to the symmetry types
\begin{equation*}%
\RectBRow{5}{1}{$k-1$}{$ $}\,,\quad\RectCRow{5}{1}{1}{$k$}{$$}{$$}, \quad \RectBRow{5}{1}{$k+1$}{$ $}
\end{equation*}
imply that the components of $W^{a(k),b}$ of these symmetry types are zero.
 The remaining component $\RectARow{5}{$k$}$ vanishes by virtue of the rank $k$ symmetric part
 of the equation
\begin{gather}%
0=D(W(k-1,1))^{a(k-1),b}+(\sigma_{(1)}{}_-^1W(k,1))^{a(k-1),b}+(\sigma_{(1)}{}_+^1W(k-2,1))^{a(k-1),b}+ \notag\\
\label{sh12}
+(\sigma_{(1)}{}_-^2W(k-1,2))^{a(k-1),b}+(\sigma_{(1)}{}_+^2W(k-1,0))^{a(k-1),b}\,.
\end{gather}
Hence,
\begin{equation}
\label{sh11}
W^{a(k),b}=0.
\end{equation}
Thereby, all fields $W^{a(k)}$ and $W^{a(k),b}$ with $k<s-1$ are shown to be zero.

\subsubsection{Dynamical equations}
\label{dynamic}
Let us now derive the dynamical equation in Minkowski and $(A)dS$ cases.
 Projection of (\ref{sh8}) with $k=s-1$ to $\RectARow{5}{$s-2$}$
 implies that  $W_{\rho}{}^{\rho a(s-2)}=0$, while the traceless
 totally symmetric part $\RectARow{5}{$s$}$ of $W^{a(s-1)}$ remains non-zero and cannot
  be gauged away.
 This is the spin $s$ dynamical field denoted $\phi_{a(s)}$. It is totally symmetric and traceless
 \begin{equation}
 \label{treacelessness1}
\phi_{a(s)}=W_{a|a(s-1)}, \quad \phi^{\rho}{}_{\rho a(s-2)}=0.
\end{equation}

 The projection of
(\ref{sh10}) to $\RectBRow{5}{1}{$k+1$}{$ $}$ expresses $W^{a(k),b}$ in terms of $dW^{a(k)}$ which is not zero
only for $k=s-1$. Analogously to Subsection 3.2.2, we can use (\ref{sh10}) and (\ref{sh12})
with $k=s-1$ to prove that $\RectBRow{5}{1}{$s$}{$ $}$ is the only non-vanishing projection
of $W^{a(s-1),b}$. One consequence of this result is that $W^{a(s-1),b}$ is traceless
  \begin{equation}
 \label{treacelessness2}
W^{\rho|}{}_{a(s-1),\rho}=0\q W^{\rho|}{}_{a(s-2)\rho,b}=0.
\end{equation}

With the help of (\ref{treacelessness1}) and (\ref{treacelessness2}), the
 contraction of indices $a$ and $\mu$ of the equation
 \begin{equation}
 \label{firstequation}
0=R_{\mu\nu|a(s-1)}=D_{\mu}\phi_{\nu a(s-1)}-D_{\nu}\phi_{\mu a(s-1)}+F_{(1)}(s-1,1)(W_{\nu|a(s-1),\mu}-W_{\mu|a(s-1),\nu})
\end{equation}
yields
 \begin{equation}
 \label{divergenceless2}
D^{\rho}\phi_{\rho a(s-1)}=0.
\end{equation}

The projection of the equation $R_{\{2\}}{}^{a(s-1),b}=0$ to the
symmetry type $\RectARow{5}{$s$}$ gives
 \begin{equation}
 \label{secondequation}
0=D^{\rho}W_{a|a(s-1),\rho}+G_{(1)}(s-1,0)(d-2)\phi_{a(s)}.
\end{equation}
Expressing $W_{a|a(s-1),\rho}$ in terms of $\phi$ by means of
(\ref{firstequation}) we obtain
\be
\label{substitution}
D^{\rho}W_{a|a(s-1),\rho}=\frac{1}{F_{(1)}(s-1,1)}(D^{\rho}D_a\phi_{\rho a(s-1)}-D^{\rho}D_{\rho}\phi_{a(s)}).
\ee
Using Eqs. (\ref{dsq}), (\ref{divergenceless2}) and  substituting
(\ref{substitution}) into (\ref{secondequation}) we obtain
 \begin{equation}
 \label{onshell}
0=D^{\rho}D_{\rho}\phi_{a(s)}-\left(\lambda^2 (d+s-2)+(d-2)H(s-1,0)\right)\phi_{a(s)},
\end{equation}
where $H(s-1,0)$ is defined in (\ref{definitionhsm}).
Using (\ref{3.39}) we finally obtain
\be
0=D^{\rho}D_{\rho}\phi_{a(s)}+\left(\lambda^2 (s^2+s(d-6)-2d+6)+m^2\right)\phi_{a(s)}.
\ee
This is the field equation for a massive spin $s$ field in the $(A)dS$ background.
The mass-like term reproduces that obtained in \cite{Higuchi}.
In the flat case  $\lambda=0$ we recover the equation (\ref{metriceq}).

It is easy to see that, the only fields that remain non-zero on shell
in the chosen gauge are the 0-forms that constitute the
 Weyl module and the projections of the 1-forms
 $W^{a(s-1),b(l)}$ to $\RectBRow{5}{3}{$s$}{$l$}$. All of them are  expressed
 by the unfolded equations via higher derivatives of the dynamical field
 $\phi_{a(s)}$.

\subsection{Spin 2 example}

Let us consider the example of a spin 2 field of generic mass, $m\ne m_t$
 (\ref{partiallymassless}) for $t=0,1$.  In this case, the full set of fields consists of 1-forms
  $W$, $W^a$, $W^{a,b}$, 0-forms
$C$, $C^a$, $C^{a,b}$ and 0-forms $C^{a(k),b(l)}$, $k\ge 2\ge l$ of the Weyl module. Unfolded equations
(\ref{finaleq1})-(\ref{finaleq3}) are
\begin{equation}
\label{1}
0=DC+f_{(0)}(1,0)e_mC^m+\alpha(0,0)W,
\end{equation}
\begin{equation}
\label{2}
0=DC^a+f_{(0)}(2,0)e_mC^{am}+g_{(0)}(0,0)e^aC+F_{(0)}(1,1)e_mC^{a,m}+\alpha(1,0)W^a,
\end{equation}
\begin{equation}
\label{3}
0=DC^{a,b}+f_{(0)}(2,1)e_m(C^{am,b}+\frac 12C^{ab,m})+G_{(0)}(1,0)(e^bC^a-e^aC^b)+\alpha(1,1)W^{a,b},
\end{equation}
\begin{equation}
\label{4}
0=DW+f_{(1)}(1,0)e_mW^m+\alpha^{-1}(0,0)f_{(0)}(1,0)F_{(0)}(1,1)e_me_nC^{m,n},
\end{equation}
\begin{equation}
\label{5}
0=DW^a+g_{(1)}(0,0)e^aW+F_{(1)}(1,1)e_mW^{a,m},
\end{equation}
\begin{equation*}
0=DW^{a,b}+G_{(1)}(1,0)(e^bW^a-e^aW^b)+2\alpha^{-1}(1,1)G_{(0)}(1,0)g_{(0)}(0,0)e^be^aC+
\end{equation*}
\begin{equation}
\label{6}
+\alpha^{-1}(1,1)(\lambda^2+G_{(0)}(1,0)F_{(0)}(1,1))(e^ae_mC^{m,b}+e^be_mC^{a,m})+
\end{equation}
\begin{equation*}
+\alpha^{-1}(1,1)f_{(0)}(2,1)F_{(0)}(2,2)e_ne_mC^{an,bm}\,
\end{equation*}
plus equations for the 0-forms in the Weyl module that follow from (\ref{1})-(\ref{6}).
The coefficients in curvatures both for 0- and 1-forms satisfy (\ref{3.39}), (\ref{normalization1}),
 (\ref{normalization2}). The gauge symmetries (\ref{gaugewc}) are
\begin{equation}
\label{7}
\delta C=-\alpha(0,0)\xi, \quad \delta C^a=-\alpha(1,0)\xi^a, \quad \delta
C^{a,b}=-\alpha(1,1)\xi^{a,b}\,,
\end{equation}
\begin{equation}
\label{8}
\delta W=D\xi+f_{(1)}(1,0)e_m\xi^m\,,
\end{equation}
\begin{equation}
\label{9}
\delta W^a=D\xi^a+g_{(1)}(0,0)e^a\xi+F_{(1)}(1,1)e_m\xi^{a,m}\,,
\end{equation}
\begin{equation}
\label{10}
\delta W^{a,b}=D\xi^{a,b}+G_{(1)}(1,0)(e^b\xi^a-e^a\xi^b).
\end{equation}
 Analogously to the massless spin 2 case, using (\ref{9}), one fixes $\xi^{a,b}$ by requiring
  the antisymmetric part of $W^{a}$
 to be zero. Then we can fix $\xi$ and $\xi^a$ setting $C=0$ and $W=0$. In this gauge
 Eq.~(\ref{1}) implies $C^m=0$. Then from (\ref{2}) it follows  that $W_{\rho|}{}^{\rho}=0$.
Eq.~(\ref{4}) states that $C^{a,b}=0$. Contracting one fiber and one space-time index of (\ref{3})
we obtain that $W_{\rho|}{}^{\rho,a}=0$. Then from (\ref{5}) it is easy to obtain
  that the dynamical field $\phi_{aa}=W_{a|a}$ is transversal. Eq. (\ref{5})
  expresses $W^{a,b}$ in terms of derivatives of $\phi_{aa}$. Plugging
 this expression  into (\ref{6}), contracting one fiber and one space-time
 index and symmetrizing the others we obtain the spin 2 equation in $AdS_d$:
 \be
 D^{\rho}D_{\rho}\phi_{aa}+(-2\lambda^2 +m^2)\phi_{aa}=0\,.
 \ee

\subsection{Massless and partially massless fields}
\label{massless}
 Let us now discuss peculiarities of the massless and partially massless cases.
 Using the type-II scaling gauge, it will be shown that
  a massive spin $s$ field
  decomposes into  the set of massless fields of spins $s'$ with
 $0\le s'\le s$  at $m=0$, $\lambda^2=0$ and
 into one  spin $s$ massless field and one spin $s-1$ massive field
 at $m=0$, $\lambda^2\neq 0$.
 Analogously, for $m=m_t$ a massive spin $s$
 field will be shown to decompose into one partially massless field of
 spin $s$ and depth $t$  and one spin $t$ massive field.

 As discussed in Section 2, massless and partially massless fields are
 characterized by the
 special values of mass $m_t$ (\ref{partiallymassless}) for which the condition
 (\ref{pmbc}) $H(k,t)=0$ holds. As follows from (\ref{definitionhsm})
 this can be achieved by setting  $F(k,t+1)=0$ and/or $G(k,t)=0$.
  Analogously, $h(t-1,l)=0$ can be achieved
 by setting $f(t,l)=0$ and/or $g(t-1,l)=0$. In the type-I scaling
gauge, from (\ref{scaling0}), (\ref{app01})-(\ref{app04}) we have
 \begin{equation*}
 F_{(0)}(k,t+1)=G_{(0)}(k,t)=f_{(0)}(t,l)=g_{(0)}(t-1,l)=
 \end{equation*}
 \begin{equation}
 \label{tI}
 =F_{(1)}(k,t+1)=G_{(1)}(k,t)=f_{(1)}(t,l)=g_{(1)}(t-1,l)=0\,.
 \end{equation}
 In the type-II scaling gauge  we have from
 (\ref{newscaling1}), (\ref{newscaling11}), (\ref{app01})-(\ref{app04})
  \begin{equation*}
 F_{(0)}(k,t+1)=G_{(0)}(k,t)=g_{(0)}(t-1,l)=0, \quad f_{(0)}(t,l)\ne 0,
 \end{equation*}
 \begin{equation}
 \label{tII}
 G_{(1)}(k,t)=f_{(1)}(t,l)=g_{(1)}(t-1,l)=0, \quad F_{(1)}(k,t+1) \ne 0.
 \end{equation}
 The main difference between the two scalings is that
 the type-II scaling gauge keeps non-zero
 $\sigma_{(0)}{}_-^1$ and $\sigma_{(1)}{}_-^2$ which is crucial
for the field equations to remain sensible for the special values of
the parameter of mass.

Indeed, for $\lambda^2=0$, the Weyl module of a spin $s'$ massless field
 consists of 0-forms valued in $\mathbf{Y}(k,s')$, $k\ge s'$.
 In Minkowski space, the unfolded massless equations
 have the form \cite{Vasiliev2001}
 \begin{equation}
 \label{weylmassless}
 R_{\{1\}}(k,s')=0=dC(k,s')+\sigma_{(0)}{}_-^1C(k+1,s'), \quad k\ge s'\,,
 \end{equation}
{\it  i.e.,} $\sigma_{(0)}{}_+^1=\sigma_{(0)}{}_+^2=\sigma_{(0)}{}_-^2=0$ and consequently
 $H_{(0)}(k,l)=h_{(0)}(k,l)=0$.
It is easy to see from (\ref{3.39}) that both $H_{(0)}(k,l)$ and $h_{(0)}(k,l)$ of the massive Weyl module
 indeed vanish at $\lambda^2=0$ and $m^2=0$. The choice of $n_{(0)}(k,l)$
 in (\ref{newscaling11})
 provides non-zero $\sigma_{(0)}{}_-^1$ in the flat massless limit.
 $H_{(0)}(k,l)=0$ leads to
   $\sigma_{(0)}{}_+^2=\sigma_{(0)}{}_-^2=0$, \ie  Eq.~(\ref{weylmassless}) is reproduced
   and the massive Weyl module decomposes into $s$ sets of 0-forms valued in $\mathbf{Y}(k,s')$
   with definite $s'<s$ and arbitrary $k\ge s$ for each set.
 Compared to the Weyl module of a massive particle, the combined set of the Weyl modules for massless fields
 of spins $s'$, $0\le s'\le s$ contains additional 0-forms that form the finite triangular
 set in Fig. 3. Let us discuss the origin of this difference in more detail.

At $\lambda^2=0$, $m^2=0$ Eq.~(\ref{3.39}) yields $H(k,l)=h(k,l)=0$ both
for 0- and for 1-forms of the triangular region of Fig. 3. In the
scaling (\ref{newscaling1}), (\ref{newscaling11}) this
 leads to vanishing of all $\sigma$
except for $\sigma_{(1)}{}_-^2$ and $\sigma_{(0)}{}_-^1$.
As a result, the triangular set of 1-forms decomposes into vertical lines,
each containing $W^{a(s'-1),b(l)}$ with $l\le s'-1$  and definite $s'<s$,
while the triangular set of 0-forms decomposes into horizontal lines,
 each containing $C^{a(k),b(s')}$ with $k\ge s'$ and definite $s'<s$.
Since from (\ref{alphaunity}) it follows that $\alpha=0$ at
$m^2=\lambda^2=0$, the curvatures for 0-forms of the triangular region do not contain 1-forms.
As a result, these 0-forms  form the completion
of the Weyl modules of the decoupled massless particles.
It can be shown that $\kappa_{++}^{12}$ and $\kappa_0$ also vanish.
$\kappa_{--}^{12}$ remains
non-zero and glues
 $W^{a(s'-1),b(s'-1)}$
  to $C^{a(s'),b(s')}$. This describes the standard gluing of gauge potentials
to the Weyl module in the massless case.

 In the case of $\lambda^2\ne 0$, a massless field is
 a particular case of $t=s-1$ of partially massless fields that allows us to
 use Fig. 2 to visualize the decomposition
 of the Weyl module in the massless limit in $(A)dS$.
 So all 0-forms of the region 5 have $s$ cells in the second row, while the 0-forms of the region 2 have $s-1$ cells in the first row. The Weyl module of a spin $s$ massive particle consists of 0-forms of regions 4 and 5.

 The $(A)dS$ Weyl module of a spin $s$ massless field
 consists of 0-forms valued in $\mathbf{Y}(k,s)$, $k\ge s$
 forming region 5 of Fig. 2.
 The unfolded equations are
 \begin{equation*}
 R_{\{1\}}(k,s)=0=dC(k,s)+\sigma_{(0)}{}_-^1C(k+1,s)+\sigma_{(0)}{}_+^1C(k-1,s), \quad k\ge s\,.
 \end{equation*}
 Decoupling of 0-forms of the Weyl module associated to a massless spin $s$
 field at $\lambda^2\ne 0$
 requires $H_{(0)}(k,s-1)=0$.
 This is indeed true for (\ref{3.39}) with $m^2=0$. The remaining 0-forms,
 valued in $\mathbf{Y}(k,l)$ with $k\ge s$, $l<s$, that form the region
 4 of Fig. 2, constitute a part
 of the Weyl module of a spin $s-1$ massive field with the specific mass
 $m_{s}$ (\ref{partiallymassless}) (recall that this value of mass
 fulfils  the condition $H_{(0)}(k,s)=h_{(0)}(s-1,l)=0$
 of the decomposition of the massive Weyl module into the
 two parts forming regions 2 and 4 of Fig. 2).

The lacking part of the Weyl module of the  spin $s-1$
 massive field contains 0-forms valued in $\mathbf{Y}(s-1,l)$ with any $l\le s-1$
 (i.e. region 2 in Fig. 2).
 In the case of $\lambda^2\ne 0$, the
1-forms $W^{a(s-1),b(l)}$ of a massless spin $s$ field decouple
from the rest 1-forms, being still glued to the massless spin $s$ Weyl module.
Since from (\ref{alphaunity}) it follows
that $\alpha(s-1,l)=0$ for $m^2=0$,
the curvatures for 0-forms $C^{a(s-1),b(l)}$ do not contain 1-forms.
These 0-forms  form a lacking part of the Weyl module of the massive
spin $s-1$ particle.
0- and 1-forms valued in $\mathbf{Y}(k,l)$
with $k<s-1$ are still mixed and form a finite triangle in Fig. 3 of a massive spin $s-1$
particle.

Let us briefly consider the partially massless case.
At $m=m_t$, the condition (\ref{pmbc}) is satisfied, thereby (\ref{tII}) is fulfilled
for the type-II scaling gauge. It is easy to see from (\ref{alphaunity}) that
\begin{equation}
\label{pmsigma3}
\alpha(k,l)=0, \qquad l\le t\le k
\end{equation}
{\it i.e.,} $\alpha(k,l)=0$ for $(k,l)$ from the region 2 of Fig. 2. It follows from
(\ref{tII}), (\ref{pmsigma3}) that unfolded equations do not mix fields of the following
two sets.

 The first set consists of mixed 1- and 0-forms of the region 1 glued to
 0-forms of regions 2 and 4 of Fig. 2. These fields describe a massive spin $t$ field of
  mass $m_s$ (\ref{partiallymassless}).

 The second set consists of 1-forms of the region 2. Since
  $\sigma_{(1)}{}_-^2(k,t+1)\ne 0$ by (\ref{tII}), these 1-forms are glued to  mixed 1- and
  0-forms valued in the region 3.  At the same time, 0- and 1-forms of the region 3 are glued to the 0-forms of the
  region 5 because $\sigma_{(0)}{}_-^1(s,l)\ne 0$.

  Compared to
  the approach of \cite{SkvortsovPM} this set provides a
  {\em non-minimal} description of a partially massless field of spin $s$
  and depth $t$
  that contains additional fields along with  constraints that express them algebraically
  in terms of derivatives of the dynamical fields modulo Stueckelberg
  degrees of freedom that are gauged away by additional Stueckelberg
  gauge symmetries. Indeed,
   to show that the two descriptions are equivalent we observe that the
  0-forms of the region 3 can be gauge fixed to zero by the
  gauge transformation (\ref{gaugewc})
  \begin{equation*}
  C^{a(k),b(l)}=0, \qquad k\le s-1,\quad l\ge t+1.
  \end{equation*}
  Then, equating curvatures (\ref{finaleq3}) for 0-forms of the region 3 to
  zero we obtain $W^{a(k),b(l)}=0$ for
  $k\le s-2$, $l\ge t+1$, while $W^{a(s-1),b(l)}$ with $l\ge t+1$ are
  expressed in terms
  of 0-forms that belong to the partially massless Weyl module
  \begin{equation}
  \label{pmreg3}
  \sigma_{(0)}{}_-^1(s,l)C(s,l)+\alpha(s-1,l)W(s-1,l)=0, \quad l\ge t+1.
  \end{equation}
  Taking into account (\ref{pmreg3}) for $l=t+1$, one finds that the term $\sigma_{(1)}{}_-^2(s-1,t+1)W(s-1,t+1)$ in the curvature $R_{\{2\}}^{a(s-1),b(t)}$
  glues  the partially massless Weyl module to the gauge
  potentials just in the same way
  as in the standard description where the gauge potentials of the region 2 are
  glued to the Weyl module via\footnote{We acknowledge a useful discussion with
  E. Skvortsov of the specificities of gluing  the
  Weyl module in the partially massless case.} \cite{Skvortsov2009nv}
  \begin{equation*}
  R_{\{2\}}{}^{a(s-1),b(t)}=e_ce_dC^{a(s-1)c,b(t)d}.
  \end{equation*}
   Thereby the non-minimal description of a partially massless
  field reduces to the standard one.
 Let us note that, although the type-II scaling gauge is most convenient
 for the analysis of the massive field decoupling in the partially
  massless limit, the standard description of a single partially massless field
  simplifies in the type-I scaling gauge.

Let us now analyze the dynamical content of the unfolded field equations
in the partially massless case where
 $\sigma_{(1)}{}_-^1(t,l)=\sigma_{(1)}{}_+^1(t-1,l)=0$ by (\ref{tII})
 and 1-forms are valued in $\mathbf{Y}(k,l)$ with $t\le k\le s-1$.
 Analogously
to the massive case we use the $\sigma_-$ cohomology analysis as in the massless case
for the set of massless fields of spins $t+1$, $t+2$, $\dots$, $s$. As a result, we are left
with a set of frame-like fields
\begin{equation}
\label{pm1}
W^{a(k)}\rightarrow\RectARow{5}{$k+1$}\oplus\RectARow{5}{$k-1$}\,, \quad t\le k \le s-1
\end{equation}
(cf (\ref{1formdecompsymm})). The leftover gauge symmetry is still given by
(\ref{gaugeleftover}) for $k$: $t\le k \le s-1$. The first component in (\ref{pm1}) of
each frame-like field except for $W^{a(s-1)}$ can be gauged away by means of a gauge
parameter $\xi^{a(k+1)}$:
\begin{equation}
\label{pmsymm}
{\cal P}_{\rm Tr=0}(W^{a|a(k)})=0\,.
\end{equation}
Here ${\cal P}_{\rm Tr=0}$ denotes the projector to the traceless part. The only gauge parameter
that cannot be fixed this way is $\xi^{a(t)}$. This is the parameter of a differential gauge symmetry.

The gauge transformation with the parameter $\xi^{a(t)}$ must be accompanied by the gauge
transformations with the parameters $\xi^{a(k)}$ with $t<k\le s-1$ to satisfy the gauge
fixing condition (\ref{pmsymm}). Indeed, the gauge transformation of ${\cal P}_{\rm Tr=0}(W^{a|a(t)})$ is
\begin{equation*}
 {\cal P}_{\rm Tr=0}(\delta W^{a|a(t)})={\cal P}_{\rm Tr=0}(D^a\xi^{a(t)})+F_{(1)}(t+1,0)\xi^{a(t+1)}
\end{equation*}
and hence
\begin{equation*}
\xi^{a(t+1)}=-\frac{{\cal P}_{\rm Tr=0}(D^a\xi^{a(t)})}{F_{(1)}(t+1,0)}.
\end{equation*}
Analogously, one can show that
\begin{equation}
\label{accomp}
\xi^{a(k)}=(-1)^{k-t}\frac{{\cal P}_{\rm Tr=0}(\overbrace{D^a\dots D^a}^{k-t}\xi^{a(t)})}{\prod_{i=t+1}^k F_{(1)}(i,0)}.
\end{equation}

 After the Stueckelberg gauge symmetry is completely fixed, we are left with
non-zero field projections
$$W^{a(k)}\rightarrow\RectARow{5}{$k-1$}\,,\quad t\le k<s-1,$$
\begin{equation}
\label{pmfields}
W^{a(s-1)}\rightarrow\RectARow{5}{$s-2$} \oplus \RectARow{5}{$s$}\,
\end{equation}
and the differential gauge symmetry (\ref{gaugeleftover}) with gauge parameters satisfying
(\ref{accomp}).

The only non-zero projection of $W^{a(t)}$ is
\begin{equation}
\label{pmdynamical1}
\psi^{a(t-1)}=W_{\rho|}{}^{\rho a(t-1)}, \qquad \delta \psi^{a(t-1)}=
D_{\rho}\xi^{\rho a(t-1)}\,.
\end{equation}
It can neither be expressed in terms of other fields nor gauged away by a
Stueckelberg gauge transformation, hence being a dynamical field.
It follows from
\begin{equation}
\label{pm3}
0=D(W(t,0))^{a(t)}+(\sigma_{(1)}{}_-^1W(t+1,0))^{a(t)}+(\sigma_{(1)}{}_-^2W(t,1))^{a(t)}
\end{equation}
that the only non-zero projections of $W^{a(t),b}$ are $\RectBRow{5}{1}{$t-1$}{$ $}$ and
$\RectARow{5}{$t$}$. The hook-type projection is expressed in terms of derivative of
$\psi$ by (\ref{pm3}). Also Eq. (\ref{pm3}) expresses the projection of $W^{a(t),b}$ to  $\RectARow{5}{$t$}$,
namely $W_{\rho|}{}^{a(t),\rho}$,
 in terms of  $D\psi$ and $W_{\rho|}{}^{\rho a(t)}$. Since there are
no other restrictions on the fields $W_{\rho|}{}^{a(t),\rho}$ and $W_{\rho|}{}^{\rho a(t)}$,
one of them, say $W_{\rho|}{}^{\rho a(t)}$, remains unrestricted and should be identified
with the second dynamical field
\begin{equation}
\label{pmdynamical2}
\varphi^{a(t)}=W_{\rho|}{}^{\rho a(t)}.
\end{equation}
Its transformation law follows from (\ref{gaugeleftover}), (\ref{accomp})
\begin{gather*}
\delta\varphi^{a(t)}=g_{(1)}(t,0)\frac{(d+2t)(d+t-2)}{d+2t-2}\xi^{a(t)}-\\
-\frac{(d+2t-2)D_{\rho}D^{\rho}\xi^{a(t)}+t (d+2t-4) D_{\rho}D^a\xi^{\rho a(t-1)}-t(t-1)D^{\rho}D^{\lambda}\xi_{\rho\lambda}{}^{a(t-2)}\eta^{aa}}{(t+1)(d+2t-2)
F_{(1)}(t+1,0)}.
\end{gather*}

The further analysis of equations for partially massless case is analogous to the massive case. $\phi^{a(s)}={\cal P}_{\rm Tr=0}(W^{a|a(s-1)})$ is the third dynamical field.
From (\ref{gaugeleftover}), (\ref{accomp}) it follows that
\begin{equation}
\label{thelastpm}
\delta\phi^{a(s)}=(-1)^{s-t-1}\frac{{\cal P}_{\rm Tr=0}(\overbrace{D^a\dots D^a}^{s-t}\xi^{a(t)})}{\prod_{i=t+1}^{s-1} F_{(1)}(i,0)}.
\end{equation}
This is in accordance with the known property
that gauge transformations of partially massless fields contain
higher derivatives \cite{Deser, ZinovievMetric, SkvortsovPM}.

\section{Action}
\label{freeaction}

The advantage of the unfolded formulation is that it operates with
gauge invariant curvatures (\ref{2.1}), (\ref{finaleq4})-(\ref{finaleq3}) which can be used to
construct gauge invariant actions. Any action, bilinear in these
curvatures, is  manifestly gauge invariant. The gauge invariant action
that describes a unitary theory should satisfy the conditions
 \begin{equation}
\label{masslessvariation3}
\frac{\delta S}{\delta W^{a(k),b(l)}}\equiv 0, \quad l\ge 2, \quad
\frac{\delta S}{\delta C^{a(q),b(r)}}\equiv 0, \quad q\ge 2,
\end{equation}
which generalize {\em the extra field decoupling condition} for
massless \cite{Vasiliev1987, LopatinVasiliev} and partially
massless \cite{SkvortsovPM} fields to the massive case.
This is equivalent to the condition that the action is free of derivatives
higher than two. Another requirement is {\em the massless decomposition condition}
which demands that the part of the action that contains two derivatives of the
dynamical fields do not mix dynamical fields associated to different  massless fields.
Here we refer to the fact, that in agreement with
construction of Zinoviev \cite{ZinovievMetric}, the massless flat limit of
a massive action should give
the sum of free massless actions for spins from $0$ to $s$. In this manner, $W^{a(k-1),b(l)}$ is
associated to the $l$-th derivative of a spin $k$ dynamical massless  field. Hence,
 such fields should not contribute to the action if $l\ge 2$.
 Analogously, in the massless case, the Weyl 0-form $C^{a(q),b(r)}$ contains
 $q$ derivatives of a spin $(q-r)$ dynamical field.
 Thereby such fields with $q\ge 2$
 also are not allowed to contribute to a unitary action.
 The extra field decoupling condition (\ref{masslessvariation3})
 along with the massless decomposition condition
 fix the action up to an overall factor and total derivatives,
 guaranteeing that it contains at most two derivatives of
 the dynamical fields and has correct structure in the massless limit.

In this section we discard
  the contribution of the 0-forms from the Weyl module to the curvatures (\ref{finaleq2}),
  (\ref{finaleq3})  using
  that this does not violate  gauge invariance because
  the Weyl tensors are gauge invariant.
  In practice, this only applies to the
curvatures $R_{\{2\}}^{a(s-1),b(s-1)}$ and $R_{\{1\}}^{a(s-1),b(s-1)}$ because
other curvatures that appear in this section
do not contain 0-forms that belong to the Weyl module.

Most of the consideration of this section
applies to an arbitrary scaling gauge (\ref{normalization1}),
(\ref{normalization2}), unless the
conditions (\ref{newscaling1}), (\ref{newscaling11}) are explicitly referred to
in the analysis of the flat massless limit.

\subsection{Action derivation}

To begin with, let us find the part of the action ${S}_{1,1}$
built from the curvatures for 1-forms,
that satisfies (\ref{masslessvariation3}) up to $C$-dependent terms.
To compensate the $C$-dependent terms we will then add
appropriate terms $S_{0,0}$ built from 0-forms so that
 \begin{equation*}
S=S_{1,1}+S_{0,0}
\end{equation*}
will satisfy (\ref{masslessvariation3}), yielding correct field equations.

${S}_{1,1}$ is constructed  in terms of the curvatures ${R}_{\{2\}}$
 analogously to the frame-like  actions for massless
 \cite{Vasiliev1987,LopatinVasiliev} and partially massless \cite{SkvortsovPM}
fields
\begin{equation}
\label{1formsgeneral}
{S}_{1,1}=\sum_{1\le l\le k\le s-1} a_{k,l}\langle {R}_{\{2\}}^{a(k),b(l)}| {R}_{\{2\}}^{a(k),b(l)}\rangle,
\end{equation}
where $\langle\dots|\dots\rangle$ is the inner product  (\ref{1formsgeneral1})
and the coefficients $a_{k,l}$ remain to be determined.

Using (\ref{conjsigma1}), the part of the variation of the action, that contains
${R}_{\{2\}}$ is
 \begin{multline}
 \label{variat}
\delta {S}_{1,1}= -2\sum_{1\le l\le k\le s-1}\Big(\langle\delta W(k,l)|\sigma_{(1)}{}_-^1{R}_{\{2\}}(k+1,l)\rangle (\frac{1}{n_{(1)}(k,l)}a_{k+1,l}-a_{k,l})+\\
+\langle\delta W(k,l)|\sigma_{(1)}{}_-^2{R}_{\{2\}}(k,l+1)\rangle (\frac{1}{N_{(1)}(k,l)}a_{k,l+1}-a_{k,l})+\\
+\langle\delta W(k,l)|\sigma_{(1)}{}_+^1{R}_{\{2\}}(k-1,l)\rangle (n_{(1)}(k-1,l)a_{k-1,l}-a_{k,l})+\\
+\langle\delta W(k,l)|\sigma_{(1)}{}_+^2{R}_{\{2\}}(k,l-1)\rangle (N_{(1)}(k,l-1)a_{k,l-1}-a_{k,l})\Big)\,.
\end{multline}
Obviously, the extra field decoupling condition (\ref{masslessvariation3})
demands
\begin{equation}
\label{afactors}
a_{k+1,l}=a_{k,l}n_{(1)}(k,l), \qquad a_{k,l+1}=a_{k,l}N_{(1)}(k,l)\,.
\end{equation}
Using (\ref{scaling0}) this gives
\begin{equation}
\label{akl0}
a_{k,l}=z(k,l)a_{s-1,1}=z(k,l)a
\end{equation}
for the type-I scaling gauge. Here $z(k,l)$ is a sign factor
\begin{equation*}
z(k,l)=\prod_{i=1}^{l-1}{\rm sign}(H(s-1,i))\prod_{j=k}^{s-2}{\rm sign}(h(j,l)).
\end{equation*}
For the type-II scaling gauge
(\ref{newscaling1}), (\ref{newscaling11}) one obtains
\begin{equation}
\label{akl}
a_{k,l}=z(k,l)a_{s-1,1}\prod_{n=1}^{l-1}{|N_{(1)}(n)|}.
\end{equation}
As explained in more detail in Section \ref{masslessaction},
to obtain non-zero variation in the flat massless limit we should set
\begin{equation}
\label{akl00}
a_{k,l}=z(k,l)a\prod_{n=0}^{l-1}{|N_{(1)}(n)|},
\end{equation}
where $a$ is both $m^2$ and $\lambda^2$ independent.

The variation (\ref{variat}) remains nonzero because the coefficients
$a_{k,l}$ are zero at  $k\le 0$. As a result,
\begin{equation}
\label{var}
\delta {S}= 2\sum_{1\le k\le s-1}a_{k,1}\left(\langle\sigma_{(1)}{}_+^2\delta W(k,0)|{R}_{\{2\}}(k,1)\rangle+
\langle\delta W(k,1)|\sigma_{(1)}{}_+^2{R}_{\{2\}}(k,0)\rangle\right)\,.
\end{equation}
Thus the extra field decoupling condition determines
${S}_{1,1}$ up to an overall factor $a$. The only terms in (\ref{var}) that contain
two derivatives of the dynamical fields are
\begin{equation*}
\langle\sigma_{(1)}{}_+^2 \delta W(k,0)|{DW}(k,1)\rangle\quad \text{and} \quad \langle\delta W(k,1)|\sigma_{(1)}{}_+^2{DW}(k,0)\rangle.
\end{equation*}
Since they do not mix fields associated to different massless fields, the massless
decomposition condition is also fulfilled.

Additional terms in the variation result from the presence of
 0-forms in the curvatures $R_{\{2\}}^{a(n),b(n)}$ (\ref{finaleq2}) valued in the rectangular Young diagrams,
{\it i.e.,} from the  terms
$
\sum_{n} a_{n,n}\langle R_{\{2\}}^{a(n),b(n)}| R_{\{2\}}^{a(n),b(n)}\rangle.
$
These give
\begin{gather}
\Delta(\delta S_{1,1})=2\Big(\sum_{1 \le n \le s-1}\big(a_{n,n}\langle\delta W(n,n)|\kappa_0 R_{\{1\}}(n,n)\rangle-
a_{n,n}\langle\delta C(n,n)|\kappa_0 R_{\{2\}}(n,n)\rangle\big)+\notag\\
\sum_{1 \le n \le s-1}\big(a_{n,n}\langle\delta W(n,n)|\kappa_{++}^{12} R_{\{1\}}(n-1,n-1)\rangle-\notag\\
\label{var1}
-a_{n-1,n-1}\frac{\alpha(n,n)}{\alpha(n-1,n-1)}\langle\delta C(n-1,n-1)|\kappa_{--}^{12} R_{\{2\}}(n,n)\rangle\big)+\\
+\sum_{1 \le n \le s-2}\big(a_{n,n}\langle\delta W(n,n)|\kappa_{--}^{12} R_{\{1\}}(n+1,n+1)\rangle-\notag\\
-a_{n+1,n+1}\frac{\alpha(n,n)}{\alpha(n+1,n+1)}\langle\delta C(n+1,n+1)|\kappa_{++}^{12} R_{\{2\}}(n,n)\rangle\big)\notag\Big),
\end{gather}
where $\alpha(k,l)$ is determined by (\ref{alphaunity0}) for the type-I scaling and by
(\ref{alphaunity}) for the type-II scaling.
Note that the terms
\begin{equation*}
\langle\delta W(s-1,s-1)|\kappa_{--}^{12} R_{\{1\}}(s,s)\rangle\quad \text{and} \quad
\langle\delta C(s,s)|\kappa_{++}^{12} R_{\{2\}}(s-1,s-1)\rangle
\end{equation*}
do not appear in (\ref{var1}) because the term $\kappa_{--}C^{a(s),b(s)}$, constructed
 by means of the 0-form that belongs to the Weyl module, have been dropped from the
 curvature $R_{\{2\}}^{a(s-1),b(s-1)}$.

The  terms (\ref{var1}) can be  compensated  by
 terms bilinear in the curvatures $R_{\{1\}}$ which have the following
 general form
\begin{gather}
\label{0formsgeneral}
\begin{split}
{S}_{0,0}=\sum_{k,l} \pi_{k,l}\int \epsilon_{l_1\dots l_d}R_{\{1\}}{}^{l_1a(k-1),b(l)}R_{\{1\}}{}^{l_2}{}_{a(k-1),b(l)}e^{l_3}\dots e^{l_d}+\\
+\sum_{k,l} \varrho_{k,l}\int \epsilon_{l_1\dots l_d}R_{\{1\}}{}^{a(k),l_1b(l-1)}R_{\{1\}}{}_{a(k)}{}^{,l_2}{}_{b(l-1)}e^{l_3}\dots e^{l_d}+\\
+\sum_{k,l} \tau_{k,l}\int \epsilon_{l_1\dots l_d}R_{\{1\}}{}^{l_1a(k),l_2b(l)}R_{\{1\}}{}_{a(k),b(l)}e^{l_3}\dots e^{l_d}+\\
+\sum_{k,l} \varphi_{k,l}\int \epsilon_{l_1\dots l_d}R_{\{1\}}{}^{l_1a(k-1),b(l)}R_{\{1\}}{}_{a(k-1)}{}^{,l_2}{}_{b(l)}e^{l_3}\dots e^{l_d}.
\end{split}
\end{gather}

It is useful to observe that almost all terms in (\ref{0formsgeneral})
 can be re-written in terms of the inner product $\langle\dots|\dots\rangle$ and some differential
  form of degree 2 built from the background vielbein. For example, the third term in
  (\ref{0formsgeneral})  is proportional to
\begin{equation}
\label{action2}
\langle R_{\{1\}}(k+1,l+1)|\sigma_{(1)}{}_+^2 \sigma_{(1)}{}_+^1R_{\{1\}}(k,l)\rangle.
\end{equation}
The analysis of terms in this form is much easier than in the general form
 (\ref{0formsgeneral}) because their
variation can be easily found  by means of Bianchi identities and
conjugation properties  (\ref{conjsigma1}).
Such a representation works  nicely provided that the
 curvatures carry enough indices. For example, to vary the term
 $$\langle\sigma_{(1)}{}_+^2 \delta W(k,l)|\sigma_{(1)}{}_-^1 \sigma_{(1)}{}_+^1 R_{\{1\}}(k,l+1)\rangle$$
 it is enough to use (\ref{conjsigma1}) for $\sigma_+^2$ to obtain
  $$-\frac{1}{N_{(1)}(k,l)}\langle \delta W(k,l)|\sigma_{(1)}{}_-^2\sigma_{(1)}{}_-^1 \sigma_{(1)}{}_+^1 R_{\{1\}}(k,l+1)\rangle.$$

 Unfortunately, such a representation does not work in the case of $l=0$ where (\ref{conjsigma1}) is
 inapplicable and the
variation of such  terms should be analyzed separately. Let these special
  terms be denoted as $S_{spec}$, while the other regular terms be denoted
   as $S_{reg}$, so that
\be
  S_{0,0}=S_{spec}+S_{reg}.
  \ee

It can be easily checked that the variation of
\begin{gather}
S_{reg}=-\sum_{reg} \langle \frac{a}{\alpha} R_{\{1\}}|\kappa R_{\{1\}}\rangle= - \sum_{n=2}^{s-1}
\left(\frac{a_{n-1,n-1}}{\alpha(n-1,n-1)}\langle R_{\{1\}}(n-1,n-1)|\kappa_{--}^{12} R_{\{1\}}(n,n)\rangle+ \right.\notag\\
 \label{action4}
+\left.\frac{a_{n,n}}{\alpha(n,n)}\langle R_{\{1\}}(n,n)|\kappa_{++}^{12} R_{\{1\}}(n-1,n-1)\rangle+\frac{a_{n,n}}{\alpha(n,n)}\langle R_{\{1\}}(n,n)|\kappa_0 R_{\{1\}}(n,n)\rangle\right)
\end{gather}
compensates (\ref{var1}) up to the following terms
\begin{equation*}
2\left(a_{1,1}\langle\delta W(1,1)|\kappa_0R_{\{1\}}
(1,1)\rangle-a_{1,1}\langle\delta C(1,1)|\kappa_0R_{\{2\}}(1,1)\rangle+\right.
\end{equation*}
\begin{equation*}
+a_{1,1}\langle\delta W(1,1)|\kappa_{++}R_{\{1\}}(0,0)\rangle+a_{1,1}\langle \kappa_{++}\delta C(0,0)|R_{\{2\}}(1,1)\rangle-
\end{equation*}
\begin{equation*}
-\frac{a_{2,1}}{\alpha(2,2)}\langle\delta C(2,1)|\sigma_{(0)}{}_-^2\kappa_{++}R_{\{1\}}(1,1)\rangle-\frac{a_{1,1}}{\alpha(1,1)}
\langle\left.\delta C(1,1)|\kappa_{--}\sigma_{(0)}{}_+^2R_{\{1\}}(2,1)\rangle\right)
\,,
\end{equation*}
which do not satisfy the extra field decoupling condition (\ref{masslessvariation3}) and
 massless decomposition condition.
Hence they
should be compensated by $S_{spec}$ of the form
\begin{equation}
\label{var2}
S_{spec}=\frac{a_{1,1}}{\alpha(1,1)}\int \epsilon_{l_1\dots l_d}\left(\mu R_{\{1\}}{}^{l_1,l_2} R_{\{1\}}+\nu R_{\{1\}}{}^{l_1,b}R_{\{1\}}{}^{l_2,}{}_{b}\right)e^{l_3}\dots e^{l_d}.
\end{equation}
 A somewhat involved calculation
shows that the proper choice of $\mu$ and $\nu$ in (\ref{var2}) is
\begin{equation}
\label{var3}
\nu=\frac{4}{(d-2)\alpha(1,1)}\left(h(1,1)\frac{d+2}{2(d-1)}-H(1,0)-\lambda^2\right), \qquad \mu=\frac{4G_{(1)}(1,0)g_{(1)}(0,0)}{\alpha(0,0)}.
\end{equation}
It can be shown that (\ref{var2}) nevertheless can be represented in the regular form
\begin{equation}
\label{var4}
S_{spec}=-\frac{a_{1,1}}{\alpha(1,1)}\langle R_{\{1\}}(1,1)|\kappa_0 R_{\{1\}}(1,1)\rangle-
\frac{2a_{1,1}}{\alpha(1,1)}\langle R_{\{1\}}(1,1)|\kappa_{++}^{12} R_{\{1\}}(0,0)\rangle.
\end{equation}

Let us note, that the conjugation relations (\ref{conjsigma1}) can be used to transform
(\ref{action4}) to the form
\begin{gather}
S_{reg}=-\sum_{reg} \langle \frac{a}{\alpha} R_{\{1\}}|\kappa R_{\{1\}}\rangle= \notag\\
\label{action41}
= \sum_{n=2}^{s-1}
\left(-\frac{a_{n,n}}{\alpha(n,n)}\langle R_{\{1\}}(n,n)|\kappa_0 R_{\{1\}}(n,n)\rangle-\frac{2a_{n,n}}{\alpha(n,n)}\langle R_{\{1\}}(n,n)|\kappa_{++}^{12} R_{\{1\}}(n-1,n-1)\rangle\right).
\end{gather}
 $S_{spec}$ (\ref{var4}) has the form of the term in brackets on the r. h. s. of (\ref{action41})  with $n=1$.

The manifestly gauge invariant action for symmetric massive fields
of any spin $s \geq 2$ is
\begin{gather}
S= \sum_{1\le l\le k\le s-1}a_{k,l}\langle R_{\{2\}}(k,l)| R_{\{2\}}(k,l)\rangle-
\notag\\
\label{finalaction}
- \sum_{n=1}^{s-1}\big(
\frac{a_{n,n}}{\alpha(n,n)}\langle R_{\{1\}}(n,n)|\kappa_0 R_{\{1\}}(n,n)\rangle+ \\
+\frac{2a_{n,n}}{\alpha(n,n)}\langle R_{\{1\}}(n,n)|\kappa_{++}^{12} R_{\{1\}}(n-1,n-1)\rangle\big)\notag.
\end{gather}

For the reader's convenience let us recall relevant notations.
The curvatures $R$ given in (\ref{finaleq4})-(\ref{finaleq3}) are contracted by means of the
scalar product $\langle\dots|\dots \rangle$ (\ref{1formsgeneral1}). The curvatures contain
operators $\sigma=\sigma_-^1+\sigma_+^1+\sigma_-^2+\sigma_+^2$ defined by (\ref{3.5})-(\ref{3.5.3})
both for 0- and 1-forms
and operators $\kappa$ given in (\ref{nonzerokappa}). The theory has the field redefinition freedom
(\ref{3.15}) independently for 0- and 1-forms.
 To fix it one should choose the scaling gauge.
 The type-I scaling gauge is
(\ref{scaling0}).
The type-II scaling gauge is (\ref{newscaling1}), (\ref{newscaling11}).
Since, for the type-II scaling, $N_{1}(k,l)$ is $k$ independent and $n_{(0)}(k,l)$ is $l$ independent,
we introduce notation $N_{(1)}(l)=N_{(1)}(k,l)$ and $n_{(0)}(k)=n_{(0)}(k,l)$. Note that $N_{(1)}(l)=-n_{(0)}(l-1)$.
The coefficients in $\sigma$ are determined by (\ref{app01})-(\ref{app04})
where $H(k,l)$ and $h(k,l)$ are given by (\ref{3.39}).
The coefficients $a_{k,l}$ and $\alpha$ are given by
(\ref{akl0}), (\ref{alphaunity0})
for the type-I scaling gauge and by (\ref{akl00}), (\ref{alphaunity})
for the type-II scaling gauge.

The variation of (\ref{finalaction}) is
\begin{equation}
\label{finalvariation}
\begin{split}
 \delta {S}= 2\sum_{1\le k\le s-1}a_{k,1}\left(\langle\sigma_{(1)}{}_+^2\delta W(k,0)|{R}_{\{2\}}(k,1)\rangle+\langle\delta W(k,1)|\sigma_{(1)}{}_+^2{R}_{\{2\}}(k,0)\rangle\right)+ \\
 +\frac{a_{1,1}\mu}{\alpha(1,1)}\int \epsilon_{l_1\dots l_d}e^{l_3}\dots e^{l_d}\big(-\alpha(0,0)\delta C^{l_1,l_2}R_{\{2\}}(0,0)-\alpha(0,0)\delta W(0,0) R_{\{1\}}^{l_1,l_2}+\\
 +(\sigma_{(0)}{}_+^2\delta C)^{l_1,l_2}R_{\{1\}}(0,0)-\delta C(0,0) (\sigma_{(0)}{}_+^2R_{\{1\}})^{l_1,l_2}\big).
 \end{split}
\end{equation}

The action (\ref{finalaction}) yields all equations used in the analysis of Section 3.3
 of the dynamical content of the unfolded field equations.
 Namely, the second term of (\ref{finalvariation}) yields the  equations that
  express the first auxiliary fields in terms of the frame-like fields
 (\ref{sh8}), (\ref{sh10}), (\ref{firstequation}). The first term yields a non-trivial equations on the
 frame-like fields (\ref{sh12}), (\ref{secondequation}). Other terms yield equations for the
  spin 0 and spin 1  fields analyzed in Section 3.3.

\subsection{Spin 2 example}
Let us consider the example of spin 2. Recall that
terms involving 0-forms from the Weyl module (namely $C^{aa,bb}$ and $C^{aa,b}$) should
 be dropped from the action (\ref{finalaction}) which has the form
 \begin{gather}
\notag
 S=a_{1,1}\langle R_{\{2\}}(1,1)|R_{\{2\}}(1,1)\rangle-\frac{a_{1,1}}{\alpha(1,1)}\langle R_{\{1\}}(1,1)|\kappa_0 R_{\{1\}}(1,1)\rangle-\\
  \label{11}
 -2
\frac{a_{1,1}}{\alpha(1,1)}\langle R_{\{1\}}(1,1)|\kappa_{++}^{12} R_{\{1\}}(0,0)\rangle.
 \end{gather}
 Its variation is
 \begin{equation}
\label{12}
\begin{split}
 \delta {S}= -2a_{1,1}\left(\langle\sigma_{(1)}{}_+^2\delta W(1,0)|{R}_{\{2\}}(1,1)\rangle-\langle\delta W(1,1)|\sigma_{(1)}{}_+^2{R}_{\{2\}}(1,0)\rangle\right)+ \\
 +\frac{a_{1,1}\mu}{\alpha(1,1)}\int \epsilon_{l_1\dots l_d}e^{l_3}\dots e^{l_d}\big(-\alpha(0,0)\delta C^{l_1,l_2}R_{\{2\}}(0,0)-\alpha(0,0)\delta W(0,0) R_{\{1\}}^{l_1,l_2}+\\
 +(\sigma_{(0)}{}_+^2\delta C)^{l_1,l_2}R_{\{1\}}(0,0)-\delta C(0,0) (\sigma_{(0)}{}_+^2R_{\{1\}})^{l_1,l_2}\big).
 \end{split}
\end{equation}
The first term gives the projection of (\ref{6}) to $\YoungB$. The second term gives (\ref{5}). Other terms yield
(\ref{1}), (\ref{4}) and traces of (\ref{2}), (\ref{3}). These are exactly
the equations used in Subsection 3.4 to derive dynamical equations.

 \subsection{Massless and partially massless limits}
\label{masslessaction}

To analyze the massless and partially limits it is convenient to
 use the type-II scaling gauge (\ref{newscaling1}), (\ref{newscaling11}). Let us fix
the coefficient $a_{s-1,1}$ in (\ref{akl}) in such a way that the product
$a_{k,1}\sigma_{(1)}{}_+^2(k,0)$ in (\ref{finalvariation}) be independent
of $m$ and $\lambda$. This insures non-zero variation
in the flat massless case. Since $\sigma_{(1)}{}_+^2(k,0)$ is proportional to $N_{(1)}{}^{-1}(0)$
(\ref{app04}), (\ref{newscaling1}), (\ref{3.39}) one
should demand
\begin{equation}
\label{normaction}
a_{s-1,1}=a |N_{(1)}(0)|,
\end{equation}
where $a$ is a coefficient independent of $m$ and $\lambda$.

For $\lambda^2\ne 0$ and $m=0$ the action still has the form (\ref{finalaction}) and its variation amounts to (\ref{finalvariation}).
As discussed in Section \ref{massless}, the 1-form fields valued in
$\mathbf{Y}(s-1,l)$ with arbitrary $l\le s-1$
decouple from the other fields. The part of the action for these fields
has the form
\begin{equation}
  \label{actionmlimit0}
 S_{m=0}= a_{s-1,l}\sum_{1\le l\le s-1}\langle R_{\{2\}}(s-1,l)| R_{\{2\}}(s-1,l)\rangle
  \end{equation}
Being formulated in terms of the 1-forms $W^{a(s-1),b(l)}$ this is the action
of \cite{LopatinVasiliev}
for a spin $s$ massless particle in $(A)dS_d$. The leftover part of (\ref{finalaction})
 is the action of a massive spin $(s-1)$ particle.

As discussed in Section \ref{massless}, at $\lambda^2=0$, 1-forms
decompose into $s$ sets, each constituted by the fields valued in $\mathbf{Y}(s'-1,l)$
with $l\le s'-1$ and various fixed $s'<s$. The action (\ref{finalaction}) also decomposes into
$s$ parts. The condition (\ref{normaction}) guarantees that the variation is
different from zero. Although the coefficients
$a_{k,l}$ that have the form (\ref{newscaling1}), (\ref{newscaling11}), (\ref{akl00})
\be
a_{k,l}=\frac{z(k,l)}{\left|\prod_{i=0}^{l-1} \big(m^2+\lambda^2(s-l-1)(s+l+d-4)\big)\right|}
\ee
 are singular in the flat massless limit $\lambda\to 0$, $m\to0$, the
 related terms in the action (\ref{finalaction})
 form a total derivative as is obvious from the fact that they do not
 contribute to the variation of the action. Dropping these terms, the action
 remains gauge invariant up to total derivative terms. However, it
cannot be written in terms of manifestly gauge invariant
curvatures for $m=\lambda=0$, giving a combination of the actions found
originally in \cite{Vasiliev1980}.

  In the partially massless limit, the action (\ref{finalaction}) also decomposes into two parts
 in agreement with the pattern described in Section \ref{massless}.
 It yields equations  for a partially massless field of spin $s$ and depth $t$ and
 for a massive spin $t$ field of mass $m_s$ (\ref{partiallymassless}).

   In the type-II scaling gauge  the partially massless part of the action
   contains the coefficients $a_{k,l}$ for $l\ge t+1$, that tend
 to infinity at $m\rightarrow m_t$ although the variation (\ref{finalvariation}) remains
 finite. Analogously to the flat massless case, the singular terms
 combine into total derivatives and can be dropped from the action.
 The remaining part of the action is non-singular
  and yields correct equations. However, it cannot be written in terms of the
  curvatures. Note that this does not contradict to the results of
  \cite{SkvortsovPM}, where the action
  for a single partially massless field is given in terms of curvatures
   because our description involves non-minimal curvatures that differ from those of
\cite{SkvortsovPM}.
  (The main difference is that $\sigma_{(1)}{}_-^2(k,t+1)$ (\ref{tII})
  is nonzero for
  the non-minimal curvatures but vanishes in the curvatures of
  \cite{SkvortsovPM}.)

\section{Conclusion}

In this paper we found the explicit form of (Stueckelberg)
gauge invariant linearized curvatures
for symmetric massive HS fields in $d$-dimensional Minkowski and $(A)dS$
space. In terms of these curvatures we constructed the manifestly gauge invariant
action as well as full unfolded field equations for free
symmetric massive HS fields.

An interesting problem for the future is to extend obtained results
to general massive mixed symmetry fields.
Note that the unfolded equations for mixed-symmetry fields have
 been recently analyzed in \cite{BIS} via radial reduction of the
 frame-like formulation of mixed symmetry fields in Minkowski space
 developed by Skvortsov \cite{SkvortsovMixed}.
 The manifestly gauge invariant action for mixed-symmetry massive
 fields in $(A)dS$ remains to be worked out.

 The challenging problem is to extend obtained results to
interacting  massive HS fields that may help to establish the correspondence
between HS gauge theory and String Theory.

\section*{Acknowledgments}

We acknowledge with gratitude the collaboration at
the early stage of this work and many useful discussions with E.~Skvortsov.
D.P. is grateful to A. Artsukevich for fruitful discussions.
This research was supported in part by,
RFBR Grant No 08-02-00963,
LSS No 1615.2008.2 and Alexander von Humboldt Foundation Grant PHYS0167
and  by Grant of UNK.

\appendix
\renewcommand{\theequation}{\Alph{section}.\arabic{equation}}
\section*{Appendix A: Conventions}
\setcounter{equation}{0}
\setcounter{section}{1}

In this paper we use the following conventions. Space-time is $d$-dimensional
           Min\-kow\-ski or $(A)dS$ space. Both base indices $\mu,\nu,\dots$
            ({\it i.e.}, indices of differential forms) and
           fiber indices $a,b,\dots$ run from $0$ to $d-1$. The fiber space is endowed with
            the mostly
           minus Minkowski metric $\eta_{ab}$. Upper and lower indices denoted by the
           same letter are contracted. Upper or lower symmetrized indices
           can be also denoted by the same letter.
           A number of symmetrized indices is often indicated in brackets by
           writing e.g. $a(k)$ instead of $(a_1\ldots a_k)$.
           In this paper we
           deal with traceless tensors that possess symmetries of two-row Young
           tableaux. Indices of the first and second rows are separated by
           comma,
           {\it i.e.}, the notation $C^{a(k),b(l)}$ automatically implies that
           \begin{equation*}
           C_{n}{}^{na(k-2),b(l)}=0,\quad C_n{}^{a(k-1),nb(l-1)}=0,\quad
           C^{a(k),}{}_n{}^{nb(l-2)}=0,\quad C^{a(k),ab(l-1)}=0.
           \end{equation*}
Sometimes instead of writing indices of tensors that possess symmetries
of a two-row Young tableau, we use the shorthand notation indicating
the lengths of rows in brackets. For instance
\begin{equation*}
C(k,l)\sim C^{a(k),b(l)}.
\end{equation*}

We often deal with similar objects for 0- and 1-forms. To distinguish
between these objects
we endow them with subscripts $(0)$ and $(1)$ which refer to the degree of a differential form
they act on. No subscript means that
the formula is the same for 0- and 1-forms.

\section*{Appendix B: Arbitrary scaling}
\setcounter{equation}{0}\setcounter{section}{2}

Here we present some useful expressions  valid for arbitrary $N_{(0)}(k,l)$, $n_{(0)}(k,l)$, $N_{(1)}(k,l)$ and $n_{(1)}(k,l)$.

   The coefficients $f_{(0)}(s,l)$ in $\sigma_{(0)}{}_-^1(s,l)$ (\ref{weylgluing}), which are responsible for
  gluing of  the Weyl module to the sector of mixed 0- and 1-forms,
   are defined by (\ref{3.11}) up to an overall factor
  \begin{equation}
  \label{weylgluing0}
  \frac{f_{(0)}(s,l)}{f_{(0)}(s,l-1)}=\frac{F_{(0)}(s,l)}{F_{(0)}(s-1,l)}\frac{s-l+1}{s-l}, \quad
  l<s
  \end{equation}
and $f_{(0)}(s,s)=0$ (\ref{fgzero}).

The compatibility condition (\ref{4.3}) with $p=0$ implies
\begin{equation}
\label{app3}
\alpha(k,l+1)=\alpha(k,l)\sqrt{\frac{N_{(1)}(k,l)}{N_{(0)}(k,l)}}, \qquad
\alpha(k+1,l)=\alpha(k,l)\sqrt{\frac{n_{(1)}(k,l)}{n_{(0)}(k,l)}}.
\end{equation}
Eq. (\ref{app3}) expresses any $\alpha(k,l)$ in terms of $\alpha(0,0)$
\begin{equation}
\label{app4}
\alpha(k,l)=\alpha(0,0)\sqrt{\frac{\prod_{i=0}^{k-1}{n_{(1)}(i,0)}\prod_{j=0}^{l-1}{ N_{(1)}(k,j)}}{\prod_{i=0}^{k-1}{n_{(0)}(i,0)}\prod_{j=0}^{l-1}{ N_{(0)}(k,j)}}}.
\end{equation}
(Note that Eqs.~(\ref{app3}) are consistent by virtue of (\ref{Nn3}).)
 Eq. (\ref{afactors}) can be used to express $a_{k,l}$ in terms of $a_{s-1,1}$:
 \begin{equation}
 \label{app5}
 a_{k,l}=a_{s-1,1}\frac{\prod_{i=1}^{l-1}N_{(1)}(s-1,i)}{\prod_{j=k}^{s-2}n_{(1)}(j,l)}.
 \end{equation}

\end{document}